\documentclass[aps,prx,twocolumn,groupedaddress,floats,showpacs,final]{revtex4-1}

\usepackage{graphicx}
\usepackage[colorlinks]{hyperref}
\usepackage{makecell}
\usepackage{mathrsfs}
\usepackage{dcolumn}

\begin{document}


\title{The halon: a quasiparticle featuring critical  charge fractionalization}


\author{Kun Chen}
\email{ckun@physics.umass.edu}
\affiliation{Department of Physics, University of Massachusetts, Amherst, Massachusetts 01003, USA}
\affiliation{Department of Physics and Astronomy, Rutgers,
The State University of New Jersey, Piscataway, NJ 08854-8019 USA}

\author{Yuan Huang}
\affiliation{Department of Physics, University of Massachusetts, Amherst, Massachusetts 01003, USA}

\author{Youjin Deng}
\affiliation{National Laboratory for Physical Sciences at Microscale and Department of Modern Physics, University of Science and Technology of China, Hefei, Anhui 230026, China}
\affiliation{Department of Physics, University of Massachusetts, Amherst, Massachusetts 01003, USA}

\author {Boris Svistunov}
\affiliation{Department of Physics, University of Massachusetts, Amherst, Massachusetts 01003, USA}
\affiliation{National Research Center ``Kurchatov Institute," 123182 Moscow, Russia }
\affiliation{Wilczek Quantum Center, School of Physics and Astronomy and T. D. Lee Institute, Shanghai Jiao Tong University, Shanghai 200240, China}
\date{\today}

\date{\today}

\begin{abstract}
The halon is a special critical state of an impurity in a quantum-critical environment. The hallmark of the halon physics is that a well-defined integer charge gets fractionalized into two parts: a microscopic core with half-integer charge and a critically large halo carrying a complementary charge of $\pm 1/2$. The halon phenomenon emerges when the impurity--environment interaction is fine-tuned to the vicinity of a boundary quantum critical point (BQCP), at which the energies of two quasiparticle states with adjacent integer charges approach each other. The universality class of such BQCP is captured by a model of pseudo-spin-$1/2$ impurity coupled to the quantum-critical environment, in such a way that the rotational symmetry in the pseudo-spin $xy$-plane is respected, with a small local ``magnetic" field  along the pseudo-spin $z$-axis playing the role of control 
parameter driving the system away from the BQCP. On the approach to BQCP, the half-integer projection of the pseudo-spin on its $z$-axis gets delocalized into a halo of critically divergent radius, capturing the essence of the phenomenon of charge fractionalization. With large-scale Monte Carlo simulations,  we confirm the existence of halons---and  quantify their universal features---in O(2) and O(3) quantum critical systems.  
\end{abstract}

\pacs{}

\maketitle


\section{Introduction}
\label{introduction}

If coupled to a ground-state many-body environment, an impurity---static or mobile, with or without internal degrees of freedom---gets dressed into a quasiparticle (polaron). The bare impurity normally carries well-defined intrinsic charges---discrete quantum numbers such as, {\rm e.g.}, the electric charge, particle number, or spin/angular momentum projection. (Some or all the intrinsic charges of the impurity can
have zero values.) The charges of the polaron, however, are not supposed to be simply inherited from those of the bare impurity. Even more importantly, the latter charges
do not have to be {\it well-defined}: In the presence of the environment charges, an impurity charge can become a bad quantum number.

The dichotomy of charge being either a good or bad quantum number is relevant not only to the properties of impurities/polarons. It applies to any kind of elementary excitations (quasiparticles). The famous example of elementary excitations characterized by well-defined particle charge  are the quasi-particles/holes
of Landau's normal Fermi liquid. Two textbook examples of quasiparticles that cannot be characterized by a well-defined particle charge are phonons in superfluids and fermionic elementary excitations in superconductors.

In what follows, we will be mostly concerned with a static impurity, in which case (the expectation value of) the impurity charge is given by the integral of the expectation value of the charge-density distortion, $\delta n ({\bf r})$, caused by the impurity (localized at the origin of the coordinates):
\begin{equation}
q = \int  \delta n ({\bf r}) \, d^dr.
\label{charge_def}
\end{equation}
When the bare impurity has its own non-zero charge, the latter should be added to the r.h.s. of Eq.~(\ref{charge_def}).
[For a mobile impurity, there is an analog of Eq.~(\ref{charge_def}), where the integrand is the environment-impurity correlator.]
The definition (\ref{charge_def}) implies a strict order of taking limits. The thermodynamic limit of infinite system size is  taken  
first. This  guarantees that $ \delta n ({\bf r})$ vanishes at $r \to \infty.$ Only then the infinite-range integration over $r$ is performed.

In the case when the charge of the dressed impurity is a good quantum number, $q$ takes on a certain integer value, which is generically insensitive to moderate changes in the coupling, $V$, between the impurity and the medium. If the impurity charge is not a good quantum number, then the value of $q$ is a continuous function of $V$, and should be understood exclusively as an expectation---rather than eigen---value.

A peculiar situation emerges in the case when the charge, while being robust with respect to small changes of the coupling $V$,  is not integer. Two characteristic examples are (i) the excitations on top of the Laughlin (fractional quantum Hall) ground state, and (ii) spinons (domain walls) in one-dimensional insulators with broken translation symmetry. For the simple reason that all the eigenvalues of the operator of the total number of particles are integer numbers, the fractional charge cannot be a good quantum number with respect to {\it all} possible measurements.
In particular, any fractional charge is doomed to come as a fluctuating integer number within a measuring protocol---nowadays experimentally achievable  with ultracold atoms in optical lattices---resulting in measuring the positions of all the particles in the system. Nevertheless, the robustness of the fractional charge in the above-mentioned examples is guaranteed by distinct topological properties of the wavefunctions and 
in this sense, does reveal universally good quantum numbers---corresponding topological charges.

In this paper, we explore a mechanism of charge fractionalization, which does not involve topological quantum numbers and is due to rather special critical conditions taking place at a boundary quantum critical point (BQCP) of the transition between two states with well-defined charges \cite{kun}. For a static impurity, a necessary condition for the effect to take place is vanishing  charge compressibility. On the other hand, a necessary condition for the absence of a trivial integer quantization of the charge is the absence of the charge gap. The two competing (!) conditions can be met in quantum-critical environments.  For an impurity without internal degrees of freedom, the BQCP is achieved by fine-tuning the coupling strength $V$ to the critical value $V_c$. The same type of BQCP emerges for a (pseudo-)spin-1/2 impurity with an appropriate coupling to the charge density of the (particle-hole symmetric) quantum-critical environment.  Here no fine-tuning is required, since the symmetry between ``spin-up" and ``spin-down" states automatically guarantees that the impurity is at the BQCP, also implying that the local ``magnetic" field---breaking the symmetry between ``spin-up" and ``spin-down" states---plays the role of the control parameter $(V-V_c)$ driving the system across the BQCP. On the approach to BQCP, the charge fractionalization comes in the form of a critically divergent halo carrying the charge $\pm1/2$.  At the transition point, there is only a microscopic core (with half-integer charge in the case of spinless impurity and a zero charge in the case of spin-1/2 impurity). Across the BQCP, the sign of the halo flips, while the structure of the core remains intact. Given the key role played by the halo, we propose the term ``halon" for this type of the critical impurity state. The numeric part of our study is based on Monte Carlo simulations with worm algorithm. 

The rest of the paper is organized as follows. In Sec.~\ref{subsec:scenarios}, we discribe possible scenarios of the effect of the environment on the impurity charge. In Sec.~\ref{subsec:halon}, we introduce the halon. In Sec.~\ref{subsec:kondo}, we discuss the relevance of the Bose Kondo model to the problem of charge quantization and the halon effect. In Sec.~\ref{subsec:rotor}, we discuss the quantum rotor model in the context of the problem of the impurity charge. The absence of charge quantization for a static impurity in a charge-compressible environment is shown in Sec.~\ref{sec:Absence}. The upper critical dimension for the halon effect is discussed  in Sec.~\ref{sec:O_N_model}. Section~\ref{sec:spin_impurity}---based on standard general considerations---establishes a number of universal relations for the halon. Section~\ref{sec:MC} deals with numeric study of the O(2) halon in 2D; it is based on worm-algorithm simulations of J-current model. In Sec.~\ref{sec:O_3}, we study a model for O(3) halon in 2D. The concluding remarks are presented in Sec.~\ref{sec:conclusion}.

\subsection{Effect of the environment on the impurity charge: Possible scenarios}
 \label{subsec:scenarios}

With respect to a certain conserved Noether charge and a {\it static} impurity, corresponding  environments fall into three categories: (i)  environments with a charge gap in the excitation spectrum, (ii) gapless environments with finite compressibility, (iii) gapless environments which is charge-incompressible. (A subtle difference between the cases of static and mobile impurity will be discussed in the end of this section.)

\begin{figure}[htbp]
 \includegraphics[scale=0.4,angle=0,width=0.7\columnwidth]{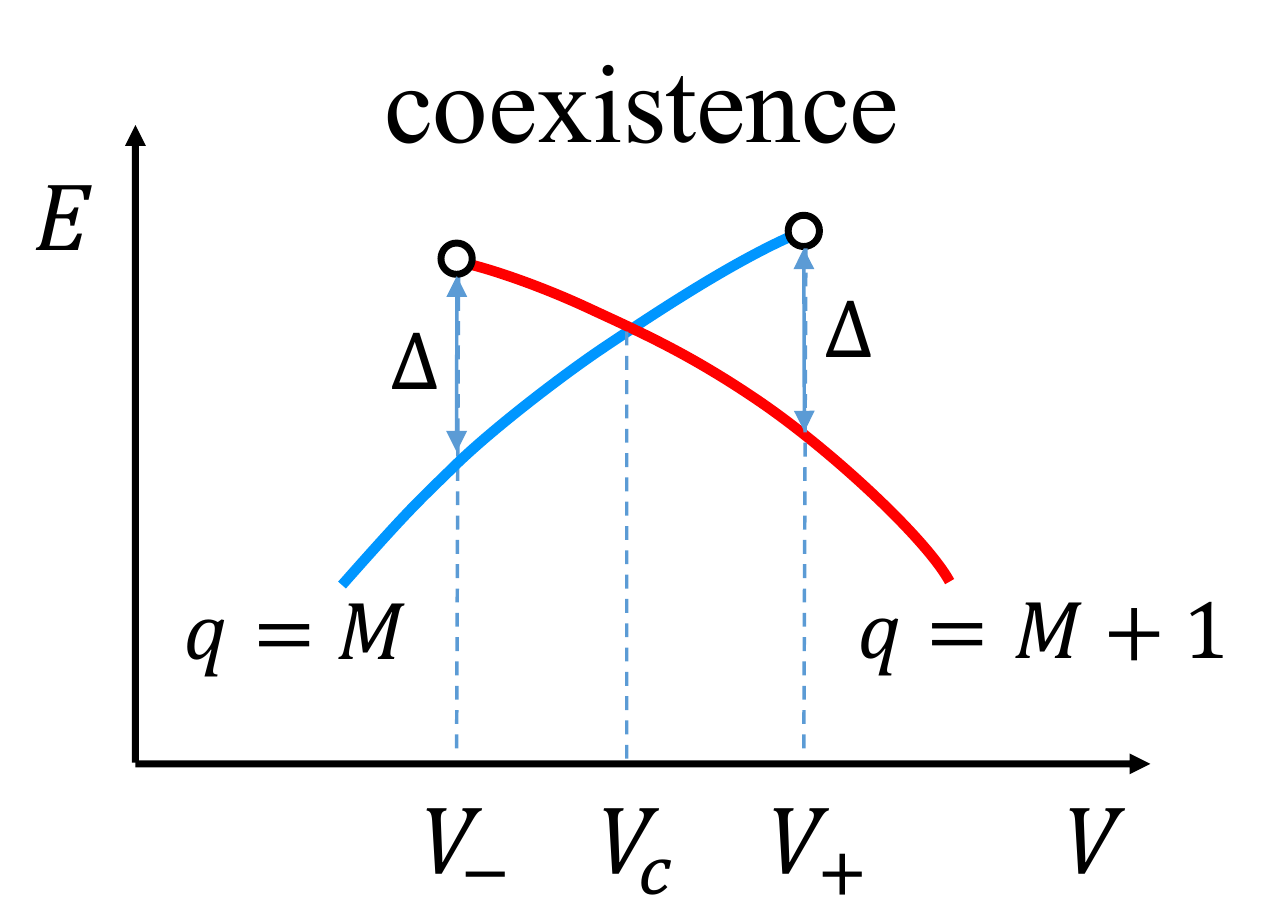}
 \centering
 \caption{Schematic behavior of impurity in a charge-gapped environment: The ground-state energy $E$ as a function of the impurity-environment interaction strength $V$, for two competing ground states. The value $V_c$ corresponds to a nominal transition between the state of the impurity charge $M$ and the state of the impurity charge $M+1$. The values $V_-$  and $V_+$ correspond to  the two end points (black dots) defined by the condition that the energy difference between the two competing states is exactly equal to the charge gap $\Delta$.}
 \label{fig:energy} 
 \end{figure}

In the presence of the charge gap, the response of the environment is quasi-perturbative. Therefore, the impurity charge is a good quantum number. Tuning the strength of the impurity-environment coupling can only cause a switch between two  states with different charges at a certain ``transition point" $V_c$. The transition, however, is merely nominal, because the charge gap protects the upper state from the decay,
as long as the energy difference between the two states remains lower than the value of the gap, $\Delta$. Hence, for each ``transition"  point $V_c$, there are two associated end points: $V_{+} > V_c$ and $V_{-} < V_c$; see Fig.~\ref{fig:energy}. Normally, the charge-gapped environment is also characterized by well-defined particle/hole elementary excitations (carrying the charge $\pm 1$) with parabolic dispersion in the long-wave-limit; Mott insulator being a very typical example. In this case, the physics of the end point is universally captured by the single-particle Schr\"odinger equation. When $V$ approaches the end point $V_{+}$ from below, the charge-$M$ impurity experiences a dramatic evolution towards a loose dimer consisting of a well-localized  charge-$(M+1)$ impurity and a weakly bound quasihole, the latter being described by the single-particle Schr\"odinger's equation with a trapping boundary condition at the origin. At the end point $V_{+}$, the quasihole unbinds. A similar picture, up to interchanging $M+1 \leftrightarrow M$ and replacing the weakly bound hole with a weakly bound particle, takes place when $V$ approaches the end point $V_{-}$ from above. 

For a static impurity in a charge-compressible environment, there is a theorem---see Sec.~\ref{sec:Absence}---stating that the charge is not quantized. The theorem stems from the fact that at $V=0$, the linear response of the charge $q$ to a small change in $V$  is proportional 
to the compressibility of the environment. If the compressibility is finite, then an arbitrarily small change in $V$ causes a change of the impurity charge,  meaning the absence of charge quantization. Of special interest is the case when the compressible environment spontaneously breaks the charge symmetry. Such an environment generates---via the anomalous averages---an effective external field mixing the states with different local charges, thus implying that a local charge cannot be a good quantum number---no matter whether the impurity is static or mobile.

\begin{figure}[htbp]
 \includegraphics[scale=1.0,angle=0,width=1.0\columnwidth]{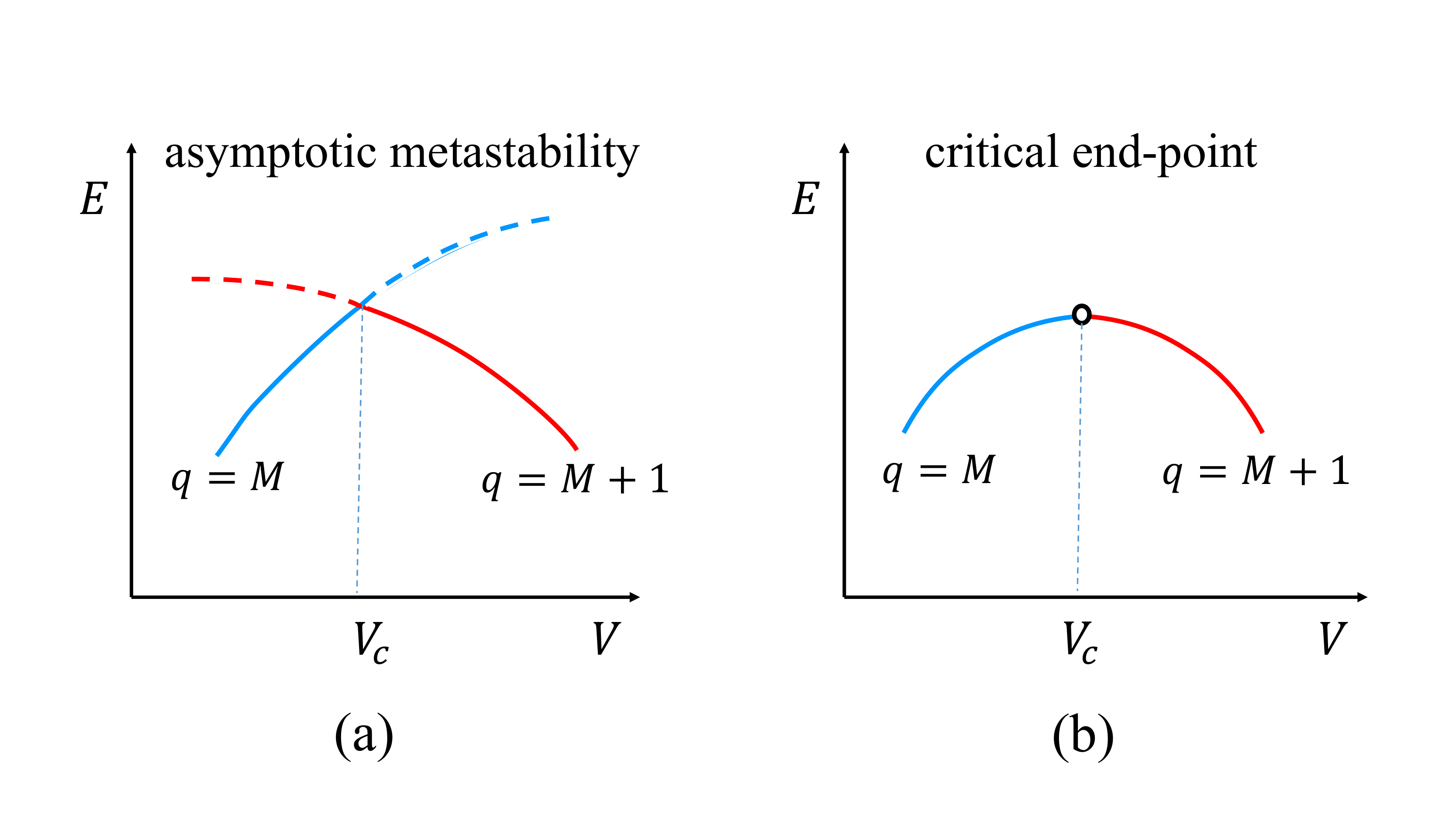}
 \centering
 \caption{Schematic behavior of impurity in a gapless,  symmetry-preserving environment in the case when the impurity charge is a good quantum number: The ground-state energy $E$ as a function of the impurity-environment coupling $V$. Two distinct scenarios are possible: (a) Two quasiparticle states with quantized charge meet at a ``first-order" phase transition point, similar to that of  Fig.~\ref{fig:energy}. The difference with the gapped environment here is that the upper states are decaying away from the critical point, the asymptotic metastability taking place in the limit $V\to V_c$. (b) Two quasiparticle states with quantized charge are connected by a continuous phase transition---a boundary quantum critical point. The notion of an excited quasiparticle state is ill defined here. }
 \label{fig:energy2} 
 \end{figure}

The case (iii) is quite rich and intriguing.  
If the charge is not quantized, there is little to discuss further. In contrast to that, the quantized charge under the conditions of the case (iii) brings about an interesting problem of the nature of the transition between two states with different charges, as a function of the coupling strength $V$. The absence of the gap
renders the upper state generically unstable with respect to the decay into the lower state. Nevertheless, the {\it asymptotic metastability} can take place in the limit $V \to V_c$ [see Fig.~\ref{fig:energy2} (a)]. Such a scenario is characteristic, for example, for the Fermi polaron \cite{fermi_polaron}. The necessary and sufficient condition for the asymptotic metastability is the vanishing (in the $V \to V_c$ limit) ratio of the decay width of the upper state to the energy difference 
between the upper and lower states. As a result, the upper state is progressively well defined---and well separated 
from the lower state---when $V$ approaches $V_c$. The critical point is thus essentially the same as in the gapped case of Fig.~\ref{fig:energy}: a doubly degenerate ground state of the impurity
at $V=V_c$.

There also exists a scenario  [see Fig.~\ref{fig:energy2} (b)]---observed in our recent numeric study \cite{kun}---when $V_c$ is the critical point of a fluctuational quantum phase transition for the impurity state,  i.e., a boundary quantum critical point. With respect to each of the two competing ground states of the impurity, the BQCP plays the role of the end point (with the halon effect taking place on the approach to it from either side). It is this scenario that will be addressed in detail in the present paper.

There is a subtle difference---with respect to the charge-quantization properties---between the cases of static and mobile impurity. For a mobile impurity, the charge can be a good quantum number even in a compressible environment, provided the charge symmetry is not broken. An important example is the normal Fermi liquid. In this environment, the charge of a static impurity is not quantized, which is guaranteed by the above-mentioned general theorem, as well as by the (system-specific) Anderson orthogonality catastrophe \cite{anderson, fermi_polaron2}. In a sharp contrast to that, a mobile impurity---the Fermi polaron---features a well-defined charge \cite{fermi_polaron,fermi_polaron2}.

Our discussion of the impurity charge in various environments will certainly be incomplete if we do
not mention the effect of trapping collapse---an infinitely large charge of a static attractive impurity---that takes place in mean-field critical environments \cite{trapping_collapse}. We refer the reader to Ref. ~\cite{trapping_collapse} for more details.

\subsection{The halon}
 \label{subsec:halon}

In the family of gapless but charge-incompressible (and symmetry-preserving) environments, of a particular interest are quantum critical systems, especially if the criticality is described by a Wilson--Fisher conformal field theory.
When an impurity is put into a quantum critical system, its interaction with the macroscopic fluctuations often leads to nontrivial emergent physics \cite{bkondo1,bkondo2,bkondo3,bkondo4,Sachdev2001,bkondo5,punk}.

\begin{figure}[h!]
 \centering
 \includegraphics[width=0.5\textwidth]{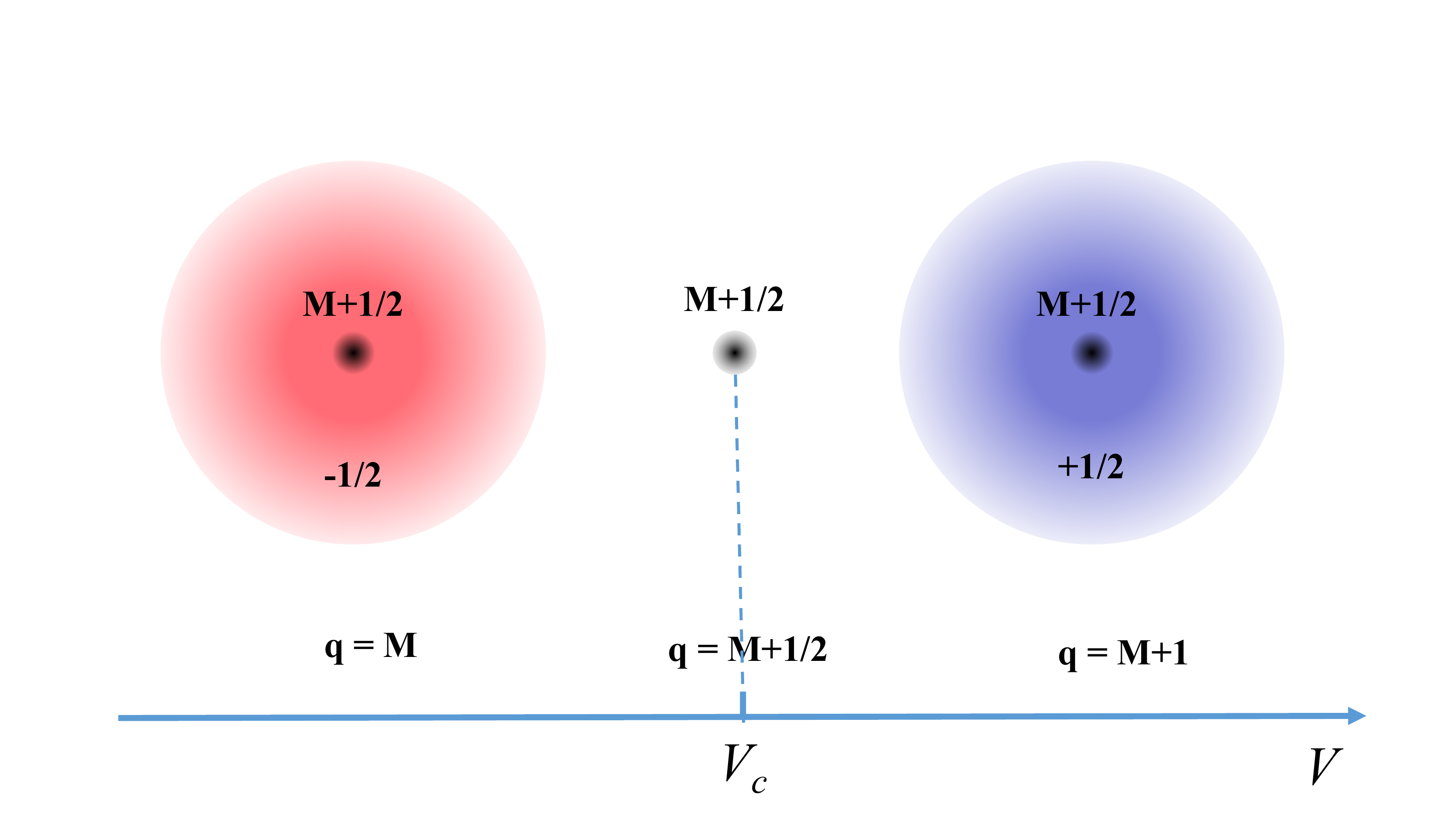}%
 \caption{Emergence of the halon. Tuning the impurity-environment interaction strength $V$ drives the quasiparticle state across a boundary quantum critical point $V_c$. the integer impurity charge separates into a short-ranged half-integer core and a large halo carrying the complementary charge of $\pm 1/2$. When $V$ crosses the transition point, the charge in the core remains the same, while the charge of the halo changes its sign, thus providing the change of $\, \pm 1$ in the net quasiparticle charge.}
 \label{halo}
 \end{figure}

Recently, we addressed the charge quantization problem for a static impurity in a two-dimensional superfluid--Mott-insulator quantum critical system \cite{kun}. Specifically, we performed worm-algorithm simulations of  the standard Bose-Hubbard model on the square lattice, with the impurity (a trapping 
center) located at the site $i=0$:
\begin{equation}
H = - \sum_{\langle ij \rangle} b_i^{\dagger} b_j + {U\over 2} \sum_i n_i(n_i -1) - \mu \sum_i n_i + V n_{i=0}.
\label{BH}
\end{equation}
Here $b_i^{\dagger}$ and $b_i$ are, respectively, bosonic creation and annihilation operators on the site $i$; the symbol $\langle \ldots \rangle$ stands for
nearest-neighbors; $U$ is the on-site interaction in units of hopping amplitude, the latter being set equal to unity. The simulations were performed at unit filling factor,  setting $U$ and the chemical potential, $\mu$, equal to their critical values, $U_c=16.7424(1)$, $\mu_c=6.21(2)$ \cite{qcp, qcp1}.  We observed a scenario of Fig.~\ref{fig:energy2} (b), with a  peculiar critical behavior of the charge distribution,  Fig.~\ref{halo}, which is likely to be generic for all fluctuational transitions between the two charge-quantized impurity states. When the center strength $V$ approaches its critical value $V_c= 6.86(8)$, the integer impurity charge separates into a short-ranged half-integer core and a large halo carrying the complementary charge of $\pm 1/2$.
The sign of the halo changes across the transition and the radius of the halo, $r_0$, diverges on the approach to $V_c$, following the critical law
\begin{equation}
r_0 \, \propto \,   |V-V_c|^{-\tilde{\nu}} , \qquad \tilde{\nu} = 2.33(5) .
\label{tilde_nu}
\end{equation}
We presented the following argument explaining why the exact half-integer quantization of the halo charge---and, correspondingly the charge of the core---follows from the very fact of existence of the halo with diverging size $r_0$. The relativistic long-range physics of the U(1) quantum criticality is particle-hole symmetric. Therefore, there should exist two halo solutions that differ only by the sign of the density distortion. In terms of these two solutions,  the net change of the impurity charge across the transition equals (plus/minus) two times the absolute value of the halo charge. Since the change of the net charge is $\pm 1$, the halo charge has to be $\pm 1/2$. More generally, the emergent particle-hole symmetry at $V \to V_c$ implies the {\it self-duality} of the boundary quantum phase transition.

One can also show \cite{kun} (corresponding analysis is rendered in our Sec.~\ref{sec:spin_impurity}) that the exponent $\tilde{\nu}$ controls some other critical properties; for example, 
the nonperturbative contribution to the energy:
\begin{equation}
E(V_c)-E(V) \,  \propto\, \left| V-V_c \right|^{\tilde{\nu}} ,
\end{equation}
as well as the power-law charge distribution in the
inner part of the halo:
\begin{equation}
|\delta n (r) |\,  \propto\,  \frac{1} {r_0^{2 - s} r^s } , \qquad s = 1 + 1/\tilde \nu \qquad (r \ll r_0) \, .
\label{inner_part}
\end{equation}

In the present paper, we perform a comprehensive study of the universal properties of the above-described BQCP.
In particular, we find the value of the critical exponent controlling the response to the local field breaking the U(1) symmetry.
We also evaluate some universal constants. Our numeric analysis is based on a spin-1/2 bosonic Kondo-type model, 
capturing the universality class of our BQCP, as explained in the next section.

\subsection{The spin-1/2 Bose Kondo model}
 \label{subsec:kondo}

Consider a model where a static (pseudo-)spin-1/2 impurity couples to the U(1) Wilson--Fisher conformal  bosonic field by the following interaction
Hamiltonian.
\begin{equation}
H_{\rm BK} = \, \gamma \, [\,{\hat S}_ +  \,\hat \psi (r = 0)\, + \, {\hat S}_ -  \,{{\hat \psi }^\dag }(r = 0)\,]\, + \, {h_z} {\hat S}_z .
\label{Bose-Kondo}
\end{equation}
Here $\hat \psi ({\bf r})$ is the field operator, ${\hat S}_z$ is the $z$-component of the operator of spin, and ${\hat S}_ -+$ and 
${\hat S}_- $ are corresponding ladder operators.

The coupling (\ref{Bose-Kondo}) preserves the global U(1) symmetry of the conformal field theory---with the global U(1) transformation involving  corresponding rotation of the spin variables in the $xy$ plane---leading to the conservation of the Noether's  charge 
\begin{equation}
Q\, = \,{{\hat S}_z} + \int d^dr\, {\hat \psi }^\dag  \hat \psi  .
\label{Bose-Kondo-charge}
\end{equation}
The U(1) Wilson--Fisher conformal  bosonic field theory is particle-hole symmetric. At $h_z=0$, the coupling (\ref{Bose-Kondo}) respects this symmetry as well, with the particle-hole transformation accompanied by flipping the spin variables with respect to the $z$-axis.

As long as we accept the concept of universality of critical phenomena, we have to conclude that the model (\ref{Bose-Kondo}) captures the universality class of the halon physics of the Hamiltonian (\ref{BH}). An advantage of the model (\ref{Bose-Kondo}) compared to (\ref{BH}) is that the particle-hole symmetry [emergent in the model (\ref{BH}) in the long-wave limit at $U=U_c$, $\mu=\mu_c$, and $V\to V_c$] now takes place at the microscopic level at $h_z=0$, meaning that the latter condition defines the halon BQCP. The correlation length of the critical response to small $h_z$,
\begin{equation}
\xi_z  \, \propto \, |h_z|^{-\nu_z}  ,
\label{xi_z}
\end{equation}
defines the halo radius:
\begin{equation}
\xi_z  \equiv r_0  , \qquad \nu_z \equiv \tilde{\nu} .
\label{xi_z__r_0}
\end{equation}
At the halon BQCP, there is yet another exponent, independent of $\tilde{\nu}$ and associated with the correlation length of the response to the local U(1)-breaking field. In terms of the model (\ref{Bose-Kondo}),  corresponding term, $h_\perp\hat{S}_x$ or $h_\perp\hat{S}_y$, is generated by coupling the spin to the magnetic field in the $xy$-plane. For the associated correlation length we then have
\begin{equation}
\xi_\perp  \, \propto \, |h_\perp|^{-\nu_\perp}  .
\label{xi_x}
\end{equation}
The other critical exponents can be related to $\nu_z$ and $\nu_\perp$ by general arguments of the theory of critical phenomena.

The model (\ref{Bose-Kondo}) is known in literature  \cite{bkondo5}, along with other models of 
impurities coupled to a bulk critical environment \cite{bkondo3,bkondo4,Sachdev2001}. However, the relevance of the model to the halon physics was realized only recently,  by Whitsitt and Sachdev in Ref.~\cite{Whitsitt_Sachdev}, and by us in the present paper.
Whitsitt and Sachdev employed the model for renormalization-group calculations, yielding, in particular, the values of the exponents $\nu_z$ and $\nu_\perp$. In what follows, we present the results of worm-algorithm simulations of a microscopic counterpart of (\ref{Bose-Kondo}) featuring the built-in $\rm U(1)\times Z_2$ symmetry at the critical point $h_z=0$. In particular, we demonstrate
(by showing that  $\nu_z = \tilde{\nu}$ within the error bars) that the spin-1/2 impurity model captures the universality class of the BQCP
of the model (\ref{BH}). We also produce a controlled numeric result for the exponent $\nu_\perp$.

Kondo-type models, where a spin-1/2 impurity couples to this or that gapless charge-conserving environment---most notably, the bosonic 
one, giving rise to the family of Bose Kondo models \cite{bkondo1,bkondo2,bkondo3,bkondo4,bkondo5}---are particularly suited (while being also interesting in a broader context) for addressing the properties of the medium with respect to the quantization of the impurity charge. In the absence of magnetic field applied to the spin-1/2 impurity: $h_z=0$, the state of the impurity  naturally corresponds to the critical point $V_c$ of the transition between two charge-quantized states (if any, see Fig.~\ref{fig:energy2}.) The absence of charge quantization corresponds to the Kondo effect, when the impurity entangles with the environment to form a singlet-type state.  The case of Fig.~\ref{fig:energy2} (a) corresponds to the absence of the Kondo effect. Here the impurity is essentially disentangled from the bulk of the system. In terms of the Kondo effect, the critical situation of Fig.~\ref{fig:energy2} (b) is marginal. 
In a sense, the effect does take place at $h_z=0$, when the spin of the impurity entangles with the whole system. However, the resulting state is
critical and thus is dramatically affected by a finite $h_z$: The characteristic length of the entanglement region---the halo---becomes finite. Note also that the $\pm 1/2$ charge of the halo naturally follows  from the first term of Eq.~(\ref{Bose-Kondo-charge}), and the self-duality of the halon  phase transition is seen at the microscopic level by the $h_z \to -h_z$ symmetry.

In the renormalization group language applied to Bose Kondo models (\ref{Bose-Kondo}), the two scenarios of Fig.~\ref{fig:energy2}, along with the case of no charge quantization, are related to the known three different types of the infrared fixed points for the running coupling constant $\gamma$ (at $h_z=0$) \cite{bose_kondo_review}. The weak-coupling fixed point $\gamma=0$  and the intermediate-coupling fixed point  $\gamma >0$ represent the cases (a) and (b), respectively, while the strong-coupling fixed point  $\gamma = \infty$ represents the case of no charge quantization.

\subsection{O($N$) quantum rotor model}
 \label{subsec:rotor}

The Bose Hubbard model (\ref{BH}), a prototypical model for the halon effect, has a disadvantage of not being particle-hole symmetric at the microscopic level. As a result, explicit upgrading the Hamiltonian (\ref{BH}) to the Bose Kondo model by replacing the last term with a spin-1/2 impurity, along the lines of Eq.~(\ref{Bose-Kondo}), has little practical sense. The halon BQCP will take place at a certain finite non-universal value of $h_z$, analogous to non-universal finite $V_c$. 

The O(2) quantum rotor model captures the universality class of the quantum phase transition of the Bose Hubbard Hamiltonian while having the desired microscopic Z$_2$ (particle-hole) symmetry. Furthermore, the quantum rotor model is naturally formulated for the group O($N$) with any $ N \geq 2$  \cite{Sachdev_book}:
\begin{equation}
\label{eq:rotor}
\hat{H}_{R}\, =\, \frac{g}{2}\sum_i \hat{\bf{L}}_i^2-J\sum_{\langle i,j \rangle} \hat{\bf n}_i\cdot \hat{\bf n}_j. 
\end{equation}
The rotor living on the site $i$ of a certain lattice is described by the operator $\hat{\bf n}_i$ of a unit vector in a 
generalized $N$-dimensional coordinate space. Corresponding momentum $\hat{\bf p}_i$ is introduced by the commutation relations
\begin{equation}
\left[\hat{n}_i^{(\alpha)}, \, \hat{p}_i^{(\beta)} \right]=i\delta_{\alpha, \beta} ,
\label{eq:mom_com}
\end{equation}
with $\alpha, \beta = 1,2, \ldots, N$ labelling the components. The rotor angular momenta $\hat{\bf{L}}_i$ are the generators of the $N$-dimensional rotational symmetry group O($N$); they have $N(N-1)/2$ independent components: 
\begin{equation}
\hat{L}_i^{(\alpha\beta)}=\hat{n}_i^{(\alpha)}\hat{p}_i^{(\beta)}-\hat{n}_i^{(\beta)}\hat{p}_i^{(\alpha)}.
\end{equation}
The first term in the Hamiltonian (\ref{eq:rotor}) is the inner product of two angular momentum operator and plays role of the kinetic energy of the rotors with $1/g$ the rotor moment of inertia, while the second term (with $J >0$) is the 
``ferromagnetic" coupling between the rotor orientations on neighboring sites. 

While elementary quantum rotors do not exist in nature, the universality of quantum critical phenomena renders the model 
(\ref{eq:rotor}) relevant to  a broad class of experimentally realizable systems. As we already mentioned, the $N=2$ case
captures the universality class of the superfluid--Mott-insulator transition in the Bose Hubbard (and similar) models. Moreover,
in the limit of large integer filling factor, the Bose Hubbard model becomes microscopically equivalent to the O(2) rotor model.
In this case,  $\hat{n}^{(1)}_j$ and $\hat{n}^{(2)}_j$ are related to the bosonic creation operator by $\hat{b}^{\dagger}_j \propto (\hat{n}^{(1)}_j + i\hat{n}^{(2)}_j)/2$. The angular momentum operator $\hat{\bf{L}}_j$ now has only one independent component $\hat{L}^{(12)}_j$. In terms of the underlying Bose Hubbard model at  large integer  filling factor, this operator corresponds to the deviation of the on-site
occupation number from its expectation value. In the case of $N=3$, the quantum rotor model describes the spin-$1/2$ ``dimerized" antiferromagnets, where each unit cell contains even number of spins. In those systems, the angular momentum $\hat{\bf{L}}_j$ has three components corresponding to three local magnetization projections.  

The type of the ground state of the model (\ref{eq:rotor}) is controlled by the dimensionless parameter $g/J$.  At $g/J \gg 1$, the first term dominates, leading the system to a quantum ``paramagnetic" state. On the other hand, the interaction term in Eq.~(\ref{eq:rotor}) is minimized by aligning the rotors. This term dominates  at $g/J \ll 1$, bringing the system to a ``magnetically" ordered state with $\langle {\bf{n}} \rangle\neq0$. The competition between the two states leads to a continuous quantum phase transition at a certain critical point $(g/J)_c$. The model (\ref{eq:rotor}) is thus representative of the O($N$) quantum criticality (known to have emergent Lorentz symmetry).
In what follows, we will be assuming that the ratio $g/J$ is kept precisely at $(g/J)_c$.

A static spinless impurity responsible for the halon effect in the quantum rotor model is introduced in a direct analogy with the Bose Hubbard model (\ref{BH}): The impurity  has to couple to a Noether charge density operator. Without loss of generality, we assume that  corresponding local field is applied to one of the components of the orbital angular momentum, say $\hat{L}^{(12)}$, on the site $i=0$:
\begin{equation}
\label{eq:rotor_imp}
\hat{H}_{R} \, \to\,  \hat{H}_{R}+V \hat{L}_{i=0}^{(12)}.
\end{equation}
This term breaks the global symmetry from O($N$) to O(2)$\times$O($N-2$), where the O(2) symmetry guarantees that the total $(12)$-component of the angular momentum, 
\begin{equation}
\label{eq:12_component}
Q=\sum_i \hat{L}^{(12)}_i ,
\end{equation}
is still a conserved charge.

Increasing the strength $V$ of the local field causes the trapped charge to change. If the charge is quantized---and our first-principle
simulations show that it is quantized at least for $N=2$ and $N=3$, boundary quantum phase transitions occur at certain {\it finite} critical values of $V$,  with a certain finite interval around the point $V=0$ corresponding to zero trapped charge. The Z$_2$ (particle-hole) symmetry of the rotor model guarantees the equivalence between the positive and negative critical values of $V_c$, but tells nothing about
the critical values themselves. A more interesting situation takes place when one introduces a spin-$1/2$ impurity [cf. Eq.~(\ref{Bose-Kondo})]:
\begin{equation}
\label{eq:spin_imp_rotor}
\hat{H}_R \to \hat{H}_R+ \gamma \left[{\hat S}_ +  \,\hat n^{(-)}_{i=0}\, + \, {\hat S}_ -  \,\hat n^{(+)}_{i=0}\right]+h_z\hat{S}_z ,
\end{equation}
where 
\begin{equation}
\hat{n}^{(\pm)}_i=(\hat{n}^{(1)}_i \pm i\hat{n}^{(2)}_i)/2
\label{ladder_op}
\end{equation}
 are the ladder operators for the $(12)$-component of the angular momentum: 
\begin{equation}
[\hat{n}^{(\pm)}_i,\,  \hat{L}^{(12)}]=\mp\hat{n}^{(\pm)}_i.
\end{equation}

In this model, the global Noether charge---the analog of (\ref{eq:12_component})---becomes
\begin{equation}
\label{eq:12_component_2}
Q=\hat{S}_z+\sum_i \hat{L}^{(12)}_i ,
\end{equation}
and the Z$_2$ (particle-hole) symmetry guarantees that $h_z=0$  is the  BQCP corresponding to the transition between the states with $Q=+1/2$ and $Q=-1/2$.

 In the long-wave-length limit, the effective field theory for the $d$-dimensional quantum rotor model at the quantum critical point is a 
 $(d+1)$-dimensional critical $\phi^4$ field theory, where $\phi$ is a continuous O($N$) vector field. The most interesting case takes place at $d=2$, where the criticality is described by a non-trivial O($N$) Wilson--Fisher conformal quantum field theory.

\section{Absence of charge quantization for a static impurity in a charge-compressible environment}
\label{sec:Absence}

Without loss of generality, we will be assuming that the charge in question is associated with the number of particles of the environment, and that we are
dealing with a continuous-space system, so that  the static impurity is represented by an external potential ${\cal U}({\bf r})$:
\begin{equation}
H_{\rm imp} = \int  {\cal U}({\bf r}) n({\bf r})\, d^d r ,
\label{ext_pot}
\end{equation}
where $n({\bf r})$ is the operator of the number density.

The statement of the section title immediately follows from the general relation
\begin{equation}
\left. {\delta q\over \delta {\cal U}({\bf r})} \right|_{{\cal U} \equiv 0} = \kappa,
\label{key_relation}
\end{equation}
where 
\begin{equation}
\kappa = {\partial n \over \partial \mu}  \qquad (\mbox{in the grand-canonical variables})
\label{kappa}
\end{equation}
is the compressibility; the partial derivative being taken in the grand-canonical variables, i.e., at fixed volume and temperature. 
We thus proceed with deriving Eq.~(\ref{key_relation}).

Consider a cubic $d$-dimensional system of linear size $L$, at a small but finite temperature $T$. The compressibility of the system is given by 
\begin{equation}
\kappa_{L} \equiv \kappa_{L}(\mu,T) = \left( {\partial n\over \partial \mu} \right)_{L,T} = \frac{\left \langle (\Delta N)^2 \right \rangle}{TL^d} .
\label{kappa_L_T}
\end{equation}
In this formula, the first equality is nothing but the definition, while the second equality---with $\left \langle (\Delta N)^2 \right \rangle$ being the variance of the total number of particles---is a straighforward implication of the definition, coming directly from the grand canonical distribution.

A subtlety arises when (\ref{kappa_L_T}) is used---in particular, in first-principle numeric simulations---to extract the thermodynamic value $\kappa_0$ of the ground-state compressibility.  Here the limits $T\to 0$ and $L\to \infty$ have to be taken under the condition of $TL^d \to \infty$, or equivalently, in the strict order (from left to right):
\begin{equation}
\kappa  = \lim_{L\to \infty} \kappa_{L} , \qquad  \quad \kappa_0 = \lim_{T\to 0} \kappa  .
\label{limits}
\end{equation}
The crucial role of the requirement 
\begin{equation}
TL^d \to \infty ,
\label{key_condition}
\end{equation}
fixing the order of limits in (\ref{limits}), is clear from the very structure of the rightmost expression in Eq.~(\ref{kappa_L_T}). The effects of finite-size quantization get completely eliminated only in 
the limit of diverging variance of the particle-number fluctuations, and the condition (\ref{key_condition}) is necessary and sufficient for this to happen.

Similarly, the definition of the impurity charge, Eq.~(\ref{charge_def}), is insensitive to the finite-size effects, if the grand canonical ensemble is used
and the ground-state thermodynamic limit is taken under the constraint  (\ref{key_condition}), or equivalently, the order of limits is exactly the same as in 
(\ref{limits}). Here the requirement (\ref{key_condition}) guarantees the absence of the spurious contributions to the integral (\ref{charge_def}) coming from the distances of the order of the system size.

Using Kubo formula, for the variational derivative in the l.h.s. of (\ref{key_relation}) we have
\begin{equation}
\left. {\delta q\over \delta {\cal U}({\bf r})} \right|_{{\cal U} \equiv 0} = \int K({\bf r}-{\bf r}', \tau) \, d^d r' d\tau
\label{Kubo}
\end{equation}
\begin{equation}
K({\bf r}, \tau) \, =\,   \langle\,  [n(0,0) - \bar{n}] \, [ n({\bf r},\tau) - \bar{n} ]\,  \rangle ,
\label{correlator_K}
\end{equation}
where $n({\bf r},\tau)$ and $\bar{n}$ are, respectively, the Matsubara operator and the expectation value of the number density. The integration over ${\bf r}'$ converts the
number density operator into the operator of the total number of particles (here we also use $\bar{n}L^d = \bar{N}$ and the fact that $N(\tau) \equiv N$, since the operator
$N$ commutes with the Hamiltonian):
\begin{equation}
\int K({\bf r}-{\bf r}', \tau) \, d^d r'\,  =\,   \langle \, [n(0,0) - \bar{n}] \,  [N - \bar{N}]\,  \rangle .
\label{correlator_K_int}
\end{equation}
The $\tau$-independence of the r.h.s. of (\ref{correlator_K_int}) trivializes the integration over $\tau$ in (\ref{Kubo}). The last simple step we need to take
to convert the r.h.s. of (\ref{Kubo}) into the r.h.s. of (\ref{kappa_L_T}) is the observation that translation invariance allows us to replace $[n(0,0) - \bar{n}]$ in
(\ref{correlator_K_int}) with $(N - \bar{N})/L^d$.

\section{Upper critical dimension for the halon effect}
\label{sec:O_N_model}

The mapping of the halon problem onto the Bose Kondo problem is very convenient for establishing the upper critical dimension for the halon effect.
Here we do it for the O($N$) quantum critical environment by rendering the known results for the spin-1/2 impurity.

Beyond the critical dimension, the physics is perturbative, see  Fig.~\ref{fig:energy2}~(a). The  condition for  this  picture to take place is
\begin{equation}
\lim_{h_z \to 0} \Gamma(h_z)/h_z \to 0 ,
\label{pert_cond}
\end{equation}
where $\Gamma(h_z)$ is the decay width of the upper branch as a function of the energy splitting $h_z > 0$.
Without loss of generality, we assume that our microscopic Hamiltonian is the quantum rotor model (\ref{eq:spin_imp_rotor}), and use
the Fermi Golden rule in the interaction picture
\begin{equation}
\label{eq:golden}
\Gamma \, = \,  \int_{-\infty}^{\infty} \langle  \hat{H}_{\rm imp}(0) \, \hat{H}_{\rm imp}(t) \rangle \,  dt ,
\end{equation}
\begin{equation}
\hat{H}_{\rm imp}(t)\, =\, e^{ih_z t} \hat{S}_{+}\,  \hat{n}^{(-)}_{i=0} (t) \, + \, e^{-ih_z t} \hat{S}_{-} \, \hat{n}^{(+)}_{i=0} (t) .
\label{H_imp_int}
\end{equation}
We thus see that $\Gamma(h_z)$ is given by the frequency-$h_z$ Fourier component of the temporal correlator (being interested in the
$h_z \to 0$ limit, we use the universal long-time asymptotic form of the correlator)
\begin{equation}
\langle \mathcal{T} \hat{n}^{(-)}_{i=0} (0)\,  \hat{n}^{(+)}_{i=0} (t) \rangle \, \propto \, 1/|t|^{d+z-2+\eta} .
\end{equation} 
Here $z=1$ is the dynamic exponent and $\eta$ is the anomalous critical exponent of the $(d+1)$-dimensional O($N$) universality class. (The exponent $\eta$ is nonzero only below the upper critical dimension $d=3$.) Performing the Fourier integral then gives
\begin{equation}
\label{eq: gamma}
\Gamma (h_z) \, \sim \, \gamma^2 h_z ^{d-2+\eta} .
\end{equation}
We conclude that the upper critical dimension for the halon physics is $d=3$, the same as for the quantum critical environment: 
Equation (\ref{eq: gamma}) becomes meaningful---consistent with (\ref{pert_cond})---at $d>3$.

While clearly characterizing the situation at $d>3$, Eq.~(\ref{eq: gamma}) leaves open the question of what happens at $d=3$. 
Here the r.h.s. of (\ref{eq: gamma}) is directly proportional to $h_z$, meaning that as long as $\gamma$ is small enough, we have
$\Gamma \ll h_z$, which seems to justify using Fermi Golden rule, but the result we get this way is not consistent with  (\ref{pert_cond})
thus calling for a more delicate analysis. 

The analysis clarifying the situation in three dimensions has been performed in Refs.~\cite{bkondo1, bkondo5, Whitsitt_Sachdev} by the
methods of perturbative renormalization group. The perturbative renormalization group treatment for a generic Bose Kondo model has been
developed in Ref.~\cite{bkondo1, bkondo2, bkondo5}. The method is based on the dimensional regularization $d=3-\epsilon$. It becomes 
controllably accurate at $\epsilon \to 0$, which is the case of our interest. The results of Refs.~\cite{bkondo1, bkondo5, Whitsitt_Sachdev}
show that the low-energy theory at the $Z_2$ symmetric point $h_z=0$ is controlled by an infrared fixed point described by a Bose Kondo impurity model with an effective coupling strength $\gamma_* \sim\epsilon$. At $\epsilon=0$, the effective coupling  $\gamma_*$ vanishes, meaning that the (renormalized) impurity and the environment are completely decoupled, which corresponds to the scenario Fig.~\ref{fig:energy2}~(a). Hence, the situation in three dimensions is essentially the same as at $d > 3$.

When $\epsilon>0$, the infrared fixed point of the perturbative RG treatment has an intermediate coupling strength $0 < \gamma_*< \infty$, suggesting a nontrivial BQCP at $d=2$. Although $\epsilon$ is not a small parameter here, the leading-order calculation seems to properly capture the qualitative physics of the system. Furthermore, comparison to our unbiased Monte Carlo results demonstrates that the third-order  $d=3-\epsilon$ calculation performed recently by Whitsitt and Sachdev \cite{Whitsitt_Sachdev} is quite accurate in reproducing the boundary quantum critical exponents.


\section{General properties of the halon}
\label{sec:spin_impurity}

In this section, we will discuss some general properties of the halon. Without loss of generality, we will be using the language of the Bose Kondo model (\ref{Bose-Kondo}) in the vicinity of the BQCP. 

Our starting point is the hyperscaling ansatz of the BQCP. As explained in the introduction, the BQCP is controlled by two independent boundary quantum critical exponents $\nu_z$ and $\nu_{\perp}$. The former is the scaling dimension of the $Z_2$-symmetry-breaking field $h_z$, the latter is the scaling dimension  of the local U(1)-symmetry-breaking field $h_{\perp}$, which is coupled to $\hat{S}_x$ (or $\hat{S}_y$) of the impurity. In the absence of dangerously irrelevant terms, one can write down the hyperscaling ansatz in the vicinity of the BQCP for the singular part of free energy,
\begin{equation}
\label{eq:hyperscaling}
F_s=b^{-z}\Phi(h_z b^{1/\nu_z}, h_{\perp} b^{1/\nu_{\perp}}, ...),
\end{equation}
for an arbitrary rescaling parameter $b$, which rescales the spatial variable as $x\rightarrow x/b$ and imaginary time variable as $\tau \rightarrow \tau/b^z$.   
Here and in what follows, we adopt a convention of omitting regular parts in critical relations.
The dynamic exponent $z>0$ is the scaling dimension of time variable of the bulk quantum criticality. In the following discussion, we will assume that $z=1$ because a bosonic system with (emergent) particle-hole symmetry normally already implies (emergent) Lorentz invariance.  However, certain edge cases (say, Lifshitz points) may  have $z>1$, and our main results can also be easily adapted. 

In the above hyperscaling ansatz, we assume that both the system size $L$ and the inverse temperature $\beta=1/T$ are much larger than the correlation length $\xi_{z,\perp}$ [as in Eq.~(\ref{xi_z}) or Eq.~(\ref{xi_x})]. Physically, it means that one needs to take the thermodynamic limit before approaching the BQCP as $h_z, h_{\perp} \rightarrow 0$.

We will consider two characteristic trajectories approaching the BQCP. One is $(h_z\rightarrow 0, h_{\perp}=0)$, and the other is $(h_z=0, h_{\perp}\rightarrow 0)$. Along these two trajectories, the characteristic length scales are $\xi_{z}$ and $\xi_{\perp}$, respectively. For each trajectory, there is only one relevant term in the hyperscaling hypothesis Eq.~(\ref{eq:hyperscaling}), so that we can choose the rescaling parameter $b$ to be the corresponding correlation length itself.

\subsection{Halon Dynamics}
\label{subsec:dynamics}

Let us discuss the universal (i.e., low-frequency) halon dynamics. Without  loss of generality, we will be using the language of the Bose Kondo model, in the Matsubara (imaginary-time or imaginary-frequency) representation. Matsubara correlation functions are directly observable in Monte Carlo simulations. Their real-frequency counterparts (linear response functions)---directly observable in the experiment---can be obtained by numeric analytic continuation.

For the spin-$1/2$ impurity in the XY Bose Kondo model, there are two independent spin correlation functions,
 \begin{equation}
\label{eq:corr}
\chi_{\alpha}(\tau)\equiv \langle \delta\hat{S}_{\alpha}(\tau) \delta\hat{S}_{\alpha}(0) \rangle,
\end{equation}
where $\alpha$ can be either longitudinal index $z$, or transverse index $\perp$, and $\delta \hat{S}_{\alpha}\equiv\hat{S}_{\alpha}-\langle \hat{S}_{\alpha}\rangle$ is the spin fluctuation operator. The two correlators yield two Matsubara (imaginary-frequency) local susceptibilities: 
 \begin{equation}
\label{eq:corr_Mats_freq}
\chi_{\alpha}(i\omega_n)=\int^{\beta}_0 \chi_{\alpha}(\tau)\,  e^{ -i\omega_n\tau} \, d\tau. 
\end{equation}

The behavior of the singular part of the static local susceptibility, $\chi_{\alpha}(i\omega_n=0)$, follows  from the hyperscaling ansatz (\ref{eq:hyperscaling}):
 \begin{equation}
\label{eq:stat_sing}
\chi_{\alpha}(i\omega_n=0)=\frac{\partial^2 F_s}{\partial h_{\alpha}^2}\sim \xi_{\alpha}^{2/\nu_{\alpha}-1}.
\end{equation}
On the approach to BQCP, the correlation length diverges as $\xi_{\alpha} \sim |h_{\alpha}|^{-\nu_{\alpha}}$, implying that that the correlation function scales as
\begin{equation}
\label{eq:corr_tau}
\chi_{\alpha}(\tau)\sim \frac{1}{\xi_{\alpha}^{2-2/\nu_{\alpha}}} \Psi_{\alpha}(\tau/ \xi_{\alpha}).
\end{equation}

Since the short-time ($\tau \ll \xi_{\alpha}$) dynamics  should be independent of the  correlation length, we conclude that
\begin{equation}
\Psi_{\alpha}(x) \sim \frac{1}{x^{2-2/\nu_{\alpha}}} ,\quad\quad\quad x\ll 1,
\label{eq:x_ll_1}
\end{equation}
or
\begin{equation}
\label{eq:corr_tau0}
\chi_{\alpha}(\tau) \sim \frac{1}{\tau^{2-2/\nu_{\alpha}}}, \quad\quad\quad \tau \ll \xi_{\alpha}.
\end{equation}
Given that the correlation function can not diverge, we get a lower bound on the critical exponents:
\begin{equation}
\label{eq:bound}
\nu_{\alpha}>1.
\end{equation}

The long-time ($\tau \gg \xi_{\alpha}$) behavior is qualitatively different.  Now the boundary-critical fluctuations are suppressed and the correlations are due exclusively to the  critical fluctuations of the environment, the only effect of the impurity being the renormalization of the global amplitude of the correlator. [In the language of the mapping onto a $(d+1)$-dimensional classical model, the  environment ``shunts" the one-dimensional ``wire" representing the impurity, imposing its correlations onto it at $\tau \gg \xi_{\alpha}$.]
We thus have (${\bf r}$ is an arbitrary spatial position appropriately  far from impurity, $r \gg \xi_\alpha$):

\begin{equation}
\label{eq:nn}
\chi_z(\tau) \propto \langle \, \hat{n}(\tau, {\bf r}) \, \hat{n}(0,{\bf r}) \, \rangle  \qquad  \quad (\tau \gg \xi_z),
\end{equation}
\begin{equation}
\label{eq:psipsi}
\chi_{\perp}(\tau) \propto \langle  \, \mathcal{T}\,  \hat{\psi} (\tau, {\bf r})\,  \hat{\psi}^{\dag} (0,{\bf r}) \, \rangle \qquad  \quad (\tau \gg \xi_{\perp}).
\end{equation}

The r.h.s of (\ref{eq:nn}) is known to demonstrate a power-law behavior $\sim 1/\tau^{2d}$, because the scaling dimension of the particle number density---or the Noether charge density---is $d$, which coincides with its canonical dimension. The  r.h.s of (\ref{eq:psipsi}) behaves as  $\sim 1/\tau^{d-1-\eta}$, where $\eta$ is the anomalous scaling dimension of the order parameter field. Therefore,
\begin{equation}
\Psi_z(x) \rightarrow {1\over x^{2d} },  \qquad  \Psi_{\perp}(x) \rightarrow {1 \over x^{d-1-\eta} } \qquad (x \gg 1).
\label{eq:x_gg_1}
\end{equation}

The experimentally relevant observable is the retarded dynamic susceptibility $\tilde{\chi}_{\alpha}(\omega)$, which depends on the real frequency $\omega$. 
The function $\tilde{\chi}_{\alpha}(\omega)$ can be obtained by analytic continuation of the Matsubara function $\chi(i\omega_n)$ to real frequencies:
\begin{equation}
\tilde{\chi}_{\alpha}(\omega)=\chi_{\alpha}(i\omega_n\rightarrow \omega+i0^+) .
\end{equation}
Using this relation in combination with Eq.~(\ref{eq:corr_tau}) yields:
\begin{equation}
\label{eq:corr_omega}
\tilde{\chi}_{\alpha}(\omega)\sim \xi_{\alpha}^{2/\nu_{\alpha}-1} \Phi_{\alpha}(\omega \xi_{\alpha}),
\end{equation}
with the universal functions $\Phi_{\alpha}$ such that 
\begin{equation}
\Phi_\alpha(x) \sim x^{1-2/\nu_{\alpha}} \quad \quad (x \gg 1), 
\end{equation}
\begin{equation}
\Phi_{z}(x) \sim x^{2d-1}, \qquad  \Phi_{\perp}(x) \sim x^{d-2-\eta} \qquad (x \ll 1) .
\end{equation}

Let us discuss the transverse dynamic susceptibility $\tilde{\chi}_{\perp}$ in more detail. 
In view of the above-mentioned shunting effect of the environment, the exponent of the function $\Psi_{\perp}(x)$ at $x \ll 1$
should be smaller or equal than the exponent of this function at $x \gg 1$. In accordance with (\ref{eq:x_ll_1}) and (\ref{eq:x_gg_1}), this implies the condition
\begin{equation}
\nu_{\perp} \, \le \, {2\over 3-d +\eta}.
\end{equation}
In the most interesting case of $d=2$, we get the following upper bound: $\nu_{\perp} \le 2/(1+\eta) <2$ (since the condition $\eta>0$ holds for any nontrivial bulk quantum criticality). Consistent with this bound,  the exponent for the O(2) bulk quantum criticality in 2D is $\nu_{\perp}=1.15(3)$ (see next section). 

\begin{figure}[h!]
 \centering
 \includegraphics[width=0.5\textwidth]{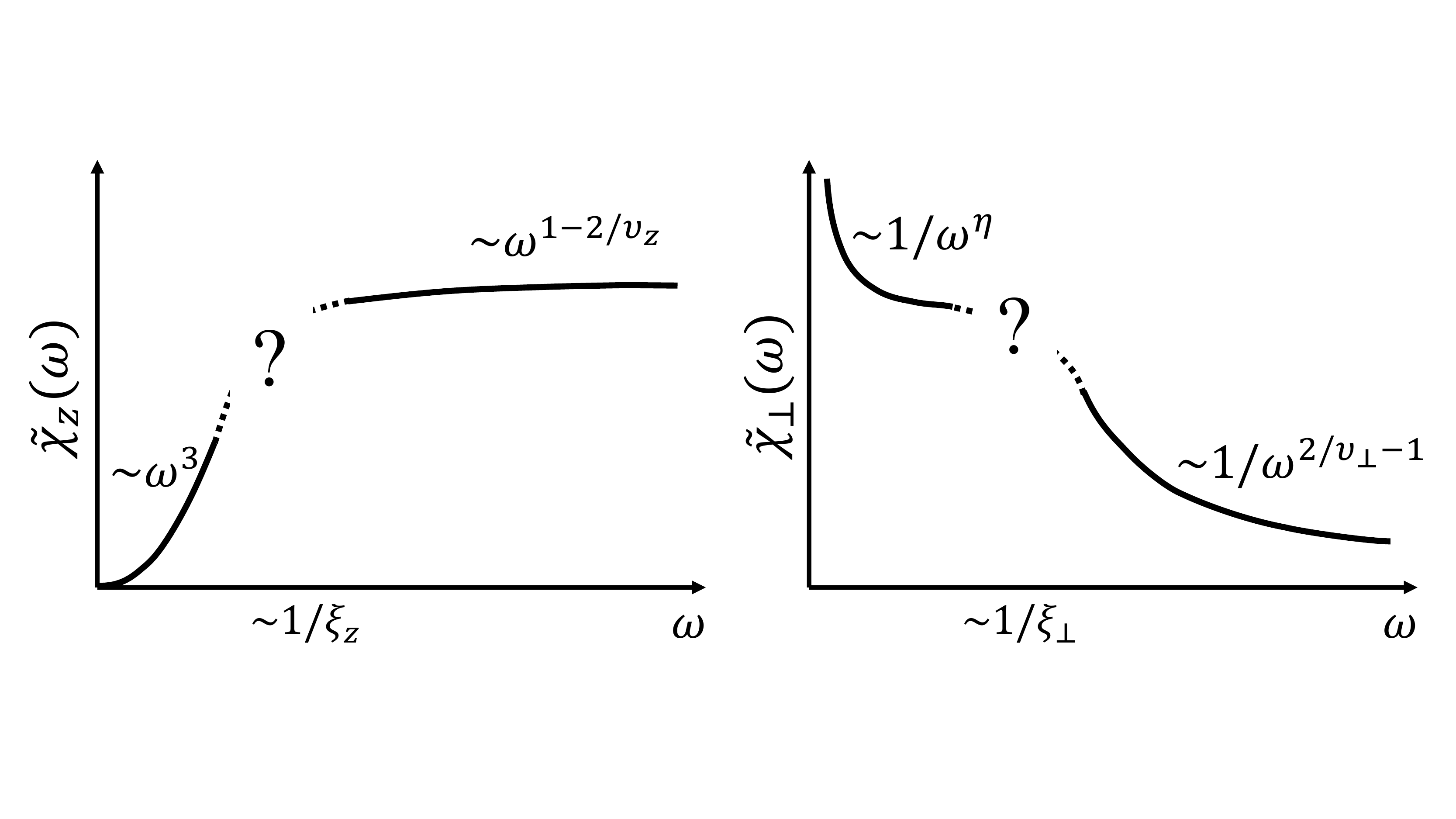}%
 \caption{\label{fig:dynamic_susp}
 The longitudinal (left) and transverse (right) dynamic susceptibilities of the impurity [for O(2) bulk quantum criticality in 2D]. The crossover from the low- to  
 high-frequency regimes takes place at $\omega \sim 1/\xi_{z,\perp}$.}
 \label{fig:halo_structure}
 \end{figure}

In Fig.~\ref{fig:dynamic_susp}, we illustrate the functions $\tilde{\chi}_z(\omega)$ and $\tilde{\chi}_{\perp}(\omega)$ for O(2) bulk quantum criticality in 2D.
The two behave very differently. The low-frequency part of $\tilde{\chi}_z(\omega)$ is $\sim \omega^3$, thus vanishing at $\omega \rightarrow 0$.
The high-frequency part of $\tilde{\chi}_z(\omega)$ behaves as $\omega^{1-2/\nu_z}$. With $\nu_z=2.33(5)>2$ \cite{kun}, we conclude that, in the high-frequency limit, $\tilde{\chi}_z(\omega)$ develops a slowly diverging pseudo plateau $~\omega^{0.14(2)}$.  In a sharp contrast with $\tilde{\chi}_z(\omega)$, the function $\tilde{\chi}_{\perp}(\omega)$  diverges at $\omega \to 0$ and vanishes at $\omega \to \infty$, and is likely to feature a shoulder at the crossover scale $\omega \sim 1/\xi_{\perp}$. The details of the crossover behavior can be established by numeric simulations. We plan to address this issue in the near future.

It is instructive to compare the dynamics of the XY Bose Kondo model to that of the SU(2) Bose Kondo model \cite{bkondo3, bkondo4}. The latter can be found in a two-dimensional O($N\ge 3$) quantum critical environment such as the paramagnetic-antiferromagnetic quantum critical point. Due to the SU(2) spin-rotational symmetry, the transverse and longitudinal dynamic susceptibilities are now essentially the same, and $\nu_z=\nu_{\perp}=\tilde{\nu}$. The low-frequency part of the susceptibility is then expected to diverge as $1/\omega^\eta$ when $\omega\rightarrow 0$, and the high-frequency part decays as $1/\omega^{2/\tilde{\nu}-1}$ as $\omega \rightarrow +\infty$, where the exponent satisfies the above-discussed inequality $\tilde{\nu}<2$. As a result, in the SU(2) Bose Kondo model, not only the transverse dynamic susceptibility but also the longitudinal one behaves as $\nu_{\perp}$ in Fig.~\ref{fig:dynamic_susp}. From this comparison we learn that the behavior of $\nu_z$  in Fig.~\ref{fig:dynamic_susp}  is a rather unique property of the XY Bose Kondo model.

\subsection{Halon Charge Density Profile}

\begin{figure}[h!]
 \centering
 \includegraphics[width=0.5\textwidth]{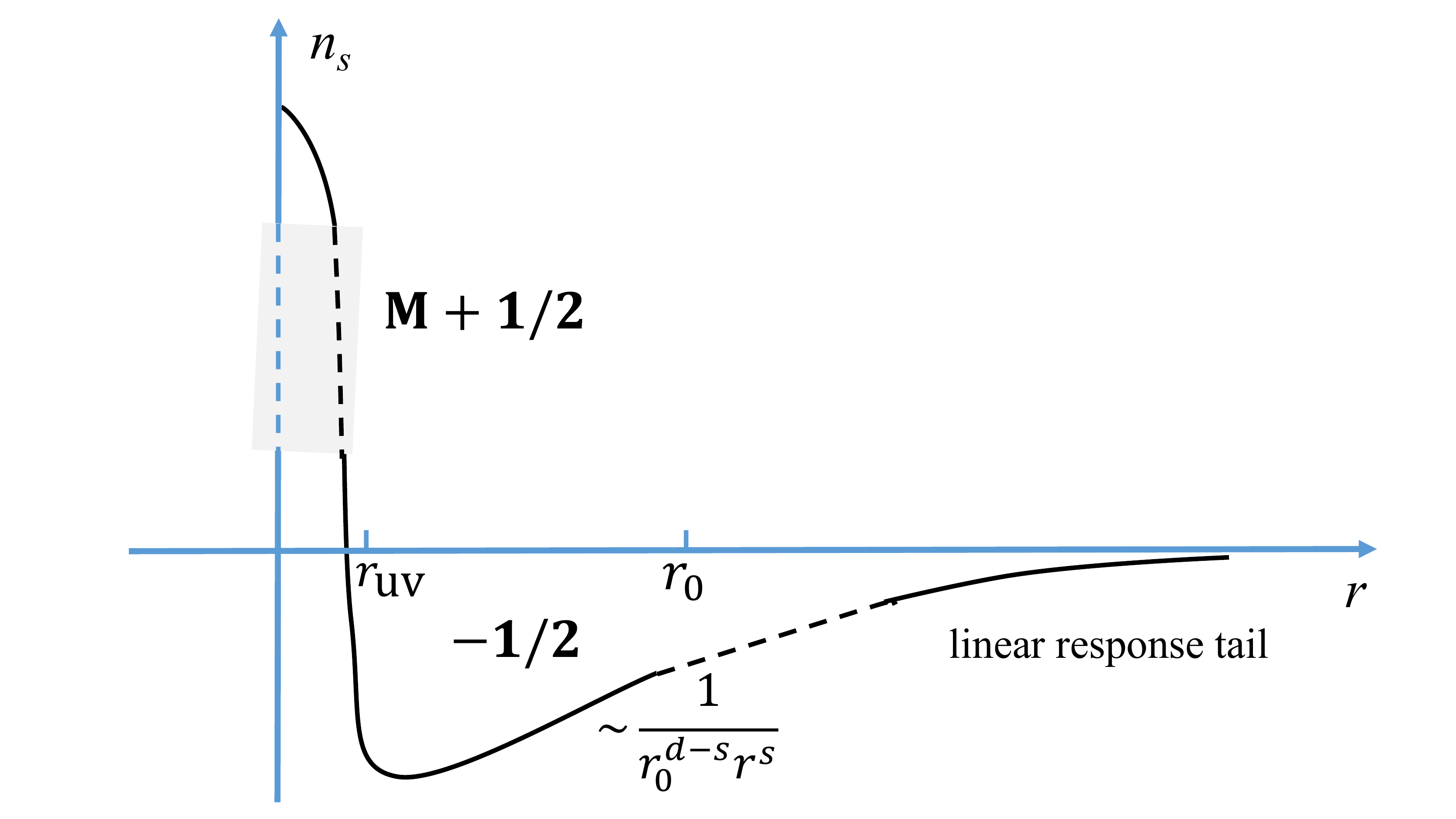}%
 \caption{The charge density profile of the halon. The charge is separated into a non-universal part and an universal part. The non-universal part is localized at a microscopic scale $r_{\rm uv}$. The universal part carries the charge $\pm 1/2$, which is distributed up to a critically large length scale $r_0$. At $r \gg r_0$, the density profile acquires a linear-response shape.}
 \label{fig:halo_structure}
 \end{figure}

Let us discuss the density profile of the halon, confining our analysis to the (most experimentally relevant) trajectory $(h_z\rightarrow 0, h_{\perp}=0)$. 
In this case, the only macroscopic characteristic length scale of the system is $\xi_z$. The physical meaning of this length scale in the spatial directions is the healing length,
\begin{equation}
r_0 \equiv \xi_z \sim  |h_z|^{-\nu_z}.
\label{eq:r_0}
\end{equation}
At $r\ll r_0$, the bosonic field is strongly influenced by the impurity and the correlation functions are controlled by the boundary quantum criticality; while at $r\gg r_0$, the impurity degrees of freedom get screened and the correlation functions of the bosonic field restore their bulk quantum critical behavior. In the halon case,  $r_0$ naturally gives the radius of the halo.

Using the standard thermodynamic relation for the averaged partial derivative of the Hamiltonian, we find the following result for the singular part
of the $z$-projection of the impurity spin (negative/positive sign corresponds to positive/negative $h_z$):
\begin{equation}
\langle S_z \rangle = {\partial F_s \over \partial h_z}    \, \propto \,  \mp  |h_z|^{\nu_z -1 }  \, \propto \,  \mp  r_0^{1/\nu_z-1}.
\label{S_z_sing}
\end{equation}
where $F_s$ is the singular part of the free energy given by Eq.~(\ref{eq:hyperscaling}). 

A crucial difference between Eqs.~(\ref{S_z_sing}) and (\ref{eq:hyperscaling}) is that the critical (singular) contribution to the free energy comes from the distances $\sim r_0$, while the $z$-projection of the impurity spin is an essentially local quantity. The latter circumstance is quite important: It implies that Eq.~(\ref{S_z_sing}) is representative of the scaling of the singular part of {\it any} generic local observable. 

In view of the divergent radius $r_0$ and scale invariance of the long-wave properties of the critical environment, the structure of the halo has to be described by a  scaling function $f$  shared by all the systems within a given universality class of the boundary quantum phase transition:
\begin{equation}
\delta n (r) = \pm  r_0^{-d} f (r/r_0) \quad \quad \quad  (r \gg r_{\rm uv}) .
\label{f_halo}
\end{equation}
Here $r_{\rm uv} \ll r_0$ is a system-specific ultraviolet cutoff.

The generic scaling for local observables, Eq.~(\ref{S_z_sing}), implies the following structure of the halo at $r\ll r_0$: 
\begin{equation}
 f(x) \propto {1\over x^s}  \qquad   \quad   (x \ll 1) ,
\label{f_halo_inner}
\end{equation}
\begin{equation}
s=d-1+1/\nu_z .
\label{exponent_s}
\end{equation}
To arrive at (\ref{f_halo_inner})--(\ref{exponent_s}), observe that by continuity, the relation $(\ref{f_halo})$ remains meaningful---at the level
of order-of-magnitude estimates---down to $r = r_{\rm uv}$, where we have $\delta n (r_{\rm uv}) \sim   |h_z|^{\nu_z -1 }  \sim   r_0^{1/\nu_z-1}$. It is also useful to write the law (\ref{f_halo_inner})--(\ref{exponent_s}) in the form 
\begin{equation}
\delta n (r)\, \propto\, {1\over r_0^{d-s} r^s} \qquad \quad (r_{\rm uv} \ll r \ll r_0) .
\label{f_halo_inner_2}
\end{equation}
Note that the above-mentioned bound $\nu_z> 1$,  implying $s < d$, guarantees the following two consistency conditions for the halo: 
(i) The integral over ${\bf r}$ should converge at $r\to 0$ and (ii) the amplitude of the singular inner part of the halo should vanish on approach to BQCP.

The shape of the outer part of the halo is dictated by a very simple perturbative physics. At $r \gg r_0$,  the boundary-critical fluctuations are suppressed leaving no room for non-linear effects. The environment thus demonstrates a generic linear response identical---up to the global amplitude---to the response caused by weak local perturbation
\begin{equation}
H_{\rm pert} = \tilde{V} \hat{n}(r=0).
\label{H_pert}
\end{equation}
Hence, at $r \gg r_0$, the charge density profile satisfies the Kubo formula
\begin{equation}
\delta n (r)\, \propto\,  \int d\tau \langle [\hat{n}({r, \tau})-\bar{n}][ \hat{n}(0, 0)-\bar{n}] \rangle .
\label{delta_n_tail_Kubo}
\end{equation}
Although in a quantum-critical environment, the particle number, which is a Noether charge, does not have anomalous scaling dimension and its density has a scaling dimension $d$. Actually, since the bulk system features the Lorentz symmetry,  the space-time density-density correlator in the r.h.s. of (\ref{delta_n_tail_Kubo}) is expected to demonstrate a power-law behavior: $\sim 1/(r^2+\tau^2)^{d}$. We thus have
\begin{equation}
 f(x) \propto {1\over x^\alpha},  \quad \alpha=2d-1 \qquad   \quad   (x \gg 1) ,
\label{f_halo_outer}
\end{equation}
or, equivalently, 
\begin{equation}
\delta n (r)\, \propto\, {r_0^{\alpha-d} \over r^{\alpha}} \qquad \quad ( r \gg r_0) .
\label{f_halo_outer_2}
\end{equation}
As we discussed in Sec. \ref{sec:Absence}, the vanishing compressibility is a necessary condition for charge quantization. In terms of the correlator (\ref{delta_n_tail_Kubo}), that means $\alpha>d$ or $d>1$. The very same condition follows from the requirement that the integral of $\delta n (r)$  over ${\bf r}$ be  convergent at $r\to \infty$. Then, according to (\ref{f_halo_outer_2}), the condition $\alpha > d$ implies the divergence of the amplitude of the linear-response tail on the approach to BQCP.

\subsection{Total Compressibility} 
\label{subsec:compressibility}

A quantum-critical system is rather fragile so that even a single halon can change substantially certain macroscopic response functions, such as, e.g., the total  compressibility. This quantity controls the response of the total number of particles to an infinitesimal change of the chemical potential. Since the total particle number is a conserved quantity, the scaling
dimension of the chemical potential is the same as that of the temperature, and the scaling ansatz for the total compressibility readily follows from the hyperscaling hypothesis, Eq.~(\ref{eq:hyperscaling}):
\begin{equation}
\label{eq:kappa}
\kappa_{\rm tot} \equiv \frac{\partial^2 F_s}{\partial \mu^2}= \frac{1}{T} C(h_z T^{-1/\nu_z}),
\end{equation}
For simplicity, we only consider the $h_z$ field. Since the temperature in experiments is finite, we take the temperature $T=1/\beta$ as the inverse characteristic length scale. In some experimental systems, like ultracold atoms in optical lattices, the system sizes $L$ can be comparable to the correlation length. Then the universal function $C(x)$ also depends on the space-time ratio $\beta/L$. We also note that the total compressibility contains no analytic contributions. This follows from the observation \cite{fisher} that a small detuning from the bulk QCP can drive the system into an incompressible insulator phase.

At the BQCP, the total compressibility $C(0)/T$ contains contribution from both the impurity and the environment. When tuning away from the BQCP, the total compressibility saturates to $C(\infty)/T$. In this limit, the impurity degree of freedom is frozen, so that the universal constant $C(\infty)$ is the same as the total compressibility of a bulk quantum critical system without the impurity. It is then natural to define the impurity compressibility at the BQCP as the difference,
\begin{equation}
\label{eq:kappa}
\kappa_{\rm imp}=\frac{C(0)-C(\infty)}{T}.
\end{equation}
If the system size is finite, the difference $C(0)-C(\infty)$ also depends on the space-time ratio $\beta/L$. Indeed, for $\beta/L=0$, which is the thermodynamic limit, Ref.~\cite{Whitsitt_Sachdev} calculates the impurity compressibility in an O(2) quantum critical environment to be roughly $0.734$; while for the ratio $\beta/L=1$, our Monte Carlo simulations determine the impurity compressibility to be $0.264(3)$ (see Sec. \ref{sec:MC}). For other space-time ratios in the range $(0,1)$, a number in between is expected for the impurity compressibility.  

In the spin language, the impurity compressibility $\kappa_{\rm imp}$ should be regarded as the effective impurity spin susceptibility. Since this quantity has a Curie-like divergence at low temperatures, similar to the SU(2) Bose Kondo problem in the Ref.~\cite{bkondo3}, one can define an effective spin $S_*$ for the impurity. The universal value of $S_*$ does not depend on the microscopic physics and can be derived from $S_*(S_*+1)/3=C(0)-C(\infty)$. In general, it is neither an integer nor a half-odd integer. 

It is  instructive to compare the behavior of $\kappa_{\rm imp}$ at the BQCP to alternative regimes of spin impurity. For the transition point of the type of Fig.~\ref{fig:energy2} (a), the spin impurity is effectively decoupled from the environment, and we expect here the physics of a free spin $\kappa_{\rm imp}= S(S+1)/3T=1/4T$ (with $S=1/2$). If the spin impurity is entirely screened by the environment, the impurity degree of freedom is no longer visible in a global response function, so that the impurity contribution $\kappa_{\rm imp}$ will be zero. Hence, measuring the total compressibility allows one to distinguish those three scenarios.

\section{Numeric simulations: 2D O(2)}
\label{sec:MC}

Simulations by worm algorithm allow us to perform a comprehensive study of the universal properties of the halon in 2D O(2) quantum critical environment. As long as we are interested in the critical properties only, we are allowed to maximally simplify the model to gain an increase in efficiency. This can be achieved by (i) working with the improved J-current model, in which the bulk finite-size corrections are almost absent and (ii) simulating the Bose Kondo model instead of the static impurity model, so that the symmetry of the both sides of BQCP 
is implemented at the microscopic level. We confirm that the Bose Kondo model (\ref{eq:spin_imp_rotor}) captures the universal critical physics of the static impurity model (\ref{eq:rotor_imp}). We also extract the values of the critical exponents and various universal constants for the halon boundary phase transition.  

\subsection{J-current model}

The simplest model for simulating the O(2) criticality is the (fully classical) J-current model ~\cite{Jcurrent}. In the $d=2+1=3$ case, the model consists of
integer currents $J$ living on the bonds of a three dimensional $L^2\times L_{\tau}$ cubic lattice, with $L$ as the size of the spatial dimensions and $L_\tau$ as the size along the ``temporal" direction (in the absence of the impurity, all the three dimensions are absolutely equivalent). The currents are subject to the zero-divergence constraint, 
\begin{equation}
{\rm div}\, J \, =\, 0
\label{divJ} ,
\end{equation}
meaning that at each site, the algebraic---incoming minus outgoing---sum of all the currents is zero. To have a really minimalistic model, one
also confines the allowed values of the bond currents to just three numbers:
\begin{equation}
J \, =\, 0, \, \pm 1.
\label{J_values} 
\end{equation}
The Hamiltonian of the model reads
\begin{equation}
\label{J-current}
 H_J \, =\,  \frac{1}{2K}  \sum_{i, \hat{e}} J_{i, \hat{e}}^2  \qquad \quad (\hat{e} = \hat{x}, \hat{y}, \hat{\tau}).
\end{equation}
Here the vector $i=(x,y,\tau)$ labels the sites on the cubic lattice by three discrete coordinates: $x$, $y$, and $z$; $\hat{x}$, $\hat{y}$, 
and $\hat{\tau}$ are the lattice unit translation vectors in corresponding directions;  $J_{i, \hat{e}} \equiv -J_{i+\hat{e}, -\hat{e}}$ is the J-current of the bond going from the site $i$ in the direction $\hat{e}$.

In terms of the mapping onto a two-dimensional system of lattice bosons (at an integer filling factor), the closed loops of currents should be understood as the worldlines of O(2) charge quanta, with $J_{i=(x,y,\tau), \hat{\tau}}$ having the meaning of the particle/hole charge on the site
$(x,y)$ at the imaginary-time moment $\tau$. The zero-divergency constraint guarantees the ``conservation of charge": The quantity 
\begin{equation}
 Q \, =\, \sum_{x,y} J_{(x,y,\tau), \hat{\tau}}
\label{J_charge}
\end{equation}
is the same for any $\tau$. This way the model (\ref{divJ})--(\ref{J-current}) describes the universal properties of the insulator-to-superfluid criticality; the corresponding transition takes place at the critical value $K_c = 0.3332052(20)$~\cite{conductivity} of the control parameter  $K$. 

Although we can simulate systems with size as large as $L=512$, many universal observables still suffer from significant non-universal finite-size corrections. To suppress those corrections caused by the renormalization of the effective coupling $K$, we find that it is very useful to introduce an improved J-current model by making the coupling strength $K$ in Eq.~(\ref{J-current}) to be current-dependent, 
\begin{equation}
\label{improved-J-current}
 H'_J \, =\,  \sum_{i, \hat{e}} \frac{1}{2K(J_{i, \hat{e}})}  J_{i, \hat{e}}^2  \qquad \quad (\hat{e} = \hat{x}, \hat{y}, \hat{\tau}).
\end{equation}
The coupling $K(J)$ needs to be fine-tuned so that the finite-size corrections caused by the renormalization of $K$ are minimized. 
In the case when there are five allowed values for the currents $(0,\pm 1,\pm2)$, we find the optimized coupling strengths to be $K(0)=K(\pm 1)=0.32944986(10)$ and $K(\pm 2)=0.16891892$ at the space-time ratio $L_\tau/L=1$.  In what follows, by default, we present the results for the minimal J-current model (\ref{J-current}). However, for some quantities with large finite-size corrections, we will employ the improved J-current model (\ref{improved-J-current}). 

We now discuss the impurity problem in the J-current models. A static spinless impurity (call it a center to avoid confusion with the spin-1/2 impurity) is introduced by the following term
\begin{equation}
\label{Jcurrent-center}
 H_{\rm center} \, =\,  V \sum_{\tau} J_{(0,0,\tau), \hat{\tau}} ,
\end{equation}
with $V$ controlling the strength of the impurity [in view of the particle-hole symmetry of the model (\ref{divJ})--(\ref{J-current}), the sign of the
impurity plays no role]. We studied the model (\ref{J-current}) with (\ref{Jcurrent-center}) in our previous work \cite{kun}, where we found that the halon 
BQCP is $V_c=1.5056(5)$, and the value of the exponent is $\tilde{\nu}=2.33(5)$.

Here we must admit that the model (\ref{divJ})--(\ref{Jcurrent-center}) is not yet optimal. Indeed, adding the term (\ref{Jcurrent-center}) to the
model (\ref{divJ})--(\ref{improved-J-current}) breaks the particle-hole symmetry of the latter, so that one has to fine-tune the parameter $V$ to the 
unknown {\it a priori} critical $V_c$. A more efficient approach is to introduce a spin-1/2 impurity thus preserving the particle-hole symmetry at zero magnetic field, analogously to how it is done for the quantum rotor model, see Eq.~(\ref{eq:spin_imp_rotor}). This amounts to
requiring that on the bonds going from the sites $(0,0,\tau)$ in the direction $\hat{\tau}$, there lives a half-integer current
\begin{equation}
S_{\tau} = \pm{1/2} .
 \label{half-current}
\end{equation}
For the sites involving the half-integer currents $S$, the zero-divergency condition also includes the algebraic sum of the two half-currents associated with this site, which guarantees the conservation of charge: the $\tau$-independence of the quantity $Q$, where
\begin{equation}
 Q \, =\, S_{\tau} + \sum_{x,y} J_{(x,y,\tau), \hat{\tau}} .
\label{J_charge_S}
\end{equation}

\begin{figure}[h!]
 \centering
 \includegraphics[width=0.35\textwidth]{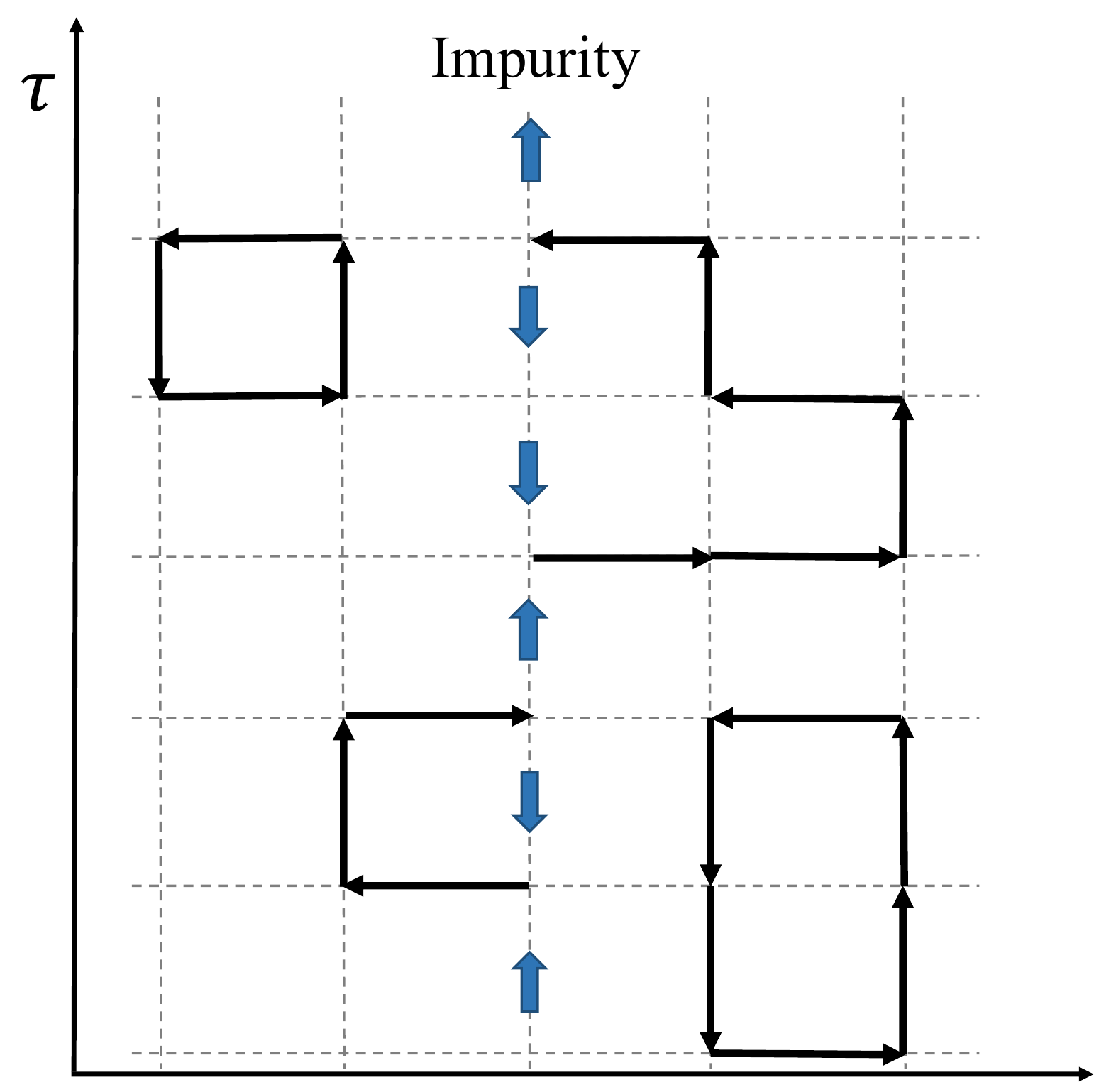}%
 \caption{Illustrative configuration of the bond currents for the J-current model with a spin-1/2  impurity at the origin in $(1+1)$ dimension. The impurity 
 half-currents $S_{\tau}$ are shown with arrows. The impurity couples to the environment via the bond currents in the spatial direction.  }
\label{fig:impurity_jcurrent}
 \end{figure}

Following the principle of minimalism, we also restrict the values of the J-currents on those bonds to zero:
\begin{equation}
J_{(0,0,\tau), \hat{\tau}} = 0 .
 \label{J_constraint}
\end{equation}
One should not confuse this constraint with decoupling the spin-1/2 impurity from the J-current environment. The coupling is still there due to the
currents $J_{(0,0,\tau), \hat{x}}$ and $J_{(0,0,\tau), \hat{y}}$; see Fig.~\ref{fig:impurity_jcurrent} for an illustration. [In terms of the analogy (a slight contrast) to the quantum rotor model (\ref{eq:spin_imp_rotor}), our spin-1/2 impurity interacts with the nearest-neighboring rotors, which allows us to apply the constraint (\ref{J_constraint}) thus safely removing the rotor on the impurity site.]

To complete the minimalistic model, we just need to introduce the coupling between the half-currents and the magnetic field:
\begin{equation}
\label{Jcurrent-imp}
 H_{\rm spin} \, =\,h_z \sum_{\tau} S_{\tau} .
\end{equation}
It is, however, important to mention that the finite-size effects prove to be rather sensitive to the strength of the coupling of the impurity to the environment. To control those effects, we change the weights of the currents $J_{(0,0,\tau), \hat{x}}$ and $J_{(0,0,\tau), \hat{y}}$ in the Hamiltonian (\ref{J-current}):
\begin{equation} 
\label{Jcurrent-imp2}
{1\over 2K} J_{(0,0,\tau),\hat{e}} ^2 \; \to \; {1\over 2K_{I}} J_{(0,0,\tau),\hat{e}} ^2  \qquad \quad (\hat{e} = \hat{x}, \hat{y}),
\end{equation}
so that while the value of $K$ is fixed at $K_c$, the value of $K_{I}$ is a free parameter, which can be optimized to improve the efficiency of the scheme.

\subsection{Global Response Functions}

Here we study the evolution of the universal compressibility on the approach to BQCP along the trajectory $(h_z \rightarrow 0, h_x=0)$. 
We set the space-time ratio to be unity ($L_\tau/L=1$), so that the scaling ansatz Eq.~(\ref{eq:kappa}) for the total compressibility predicts $\kappa_{\rm tot}/L \sim C(h_z L^{1/\nu_z})$, where $C(x)$ is a universal function. 

\begin{figure}[h!]
\centering
  \centering
  \includegraphics[width=1.\linewidth]{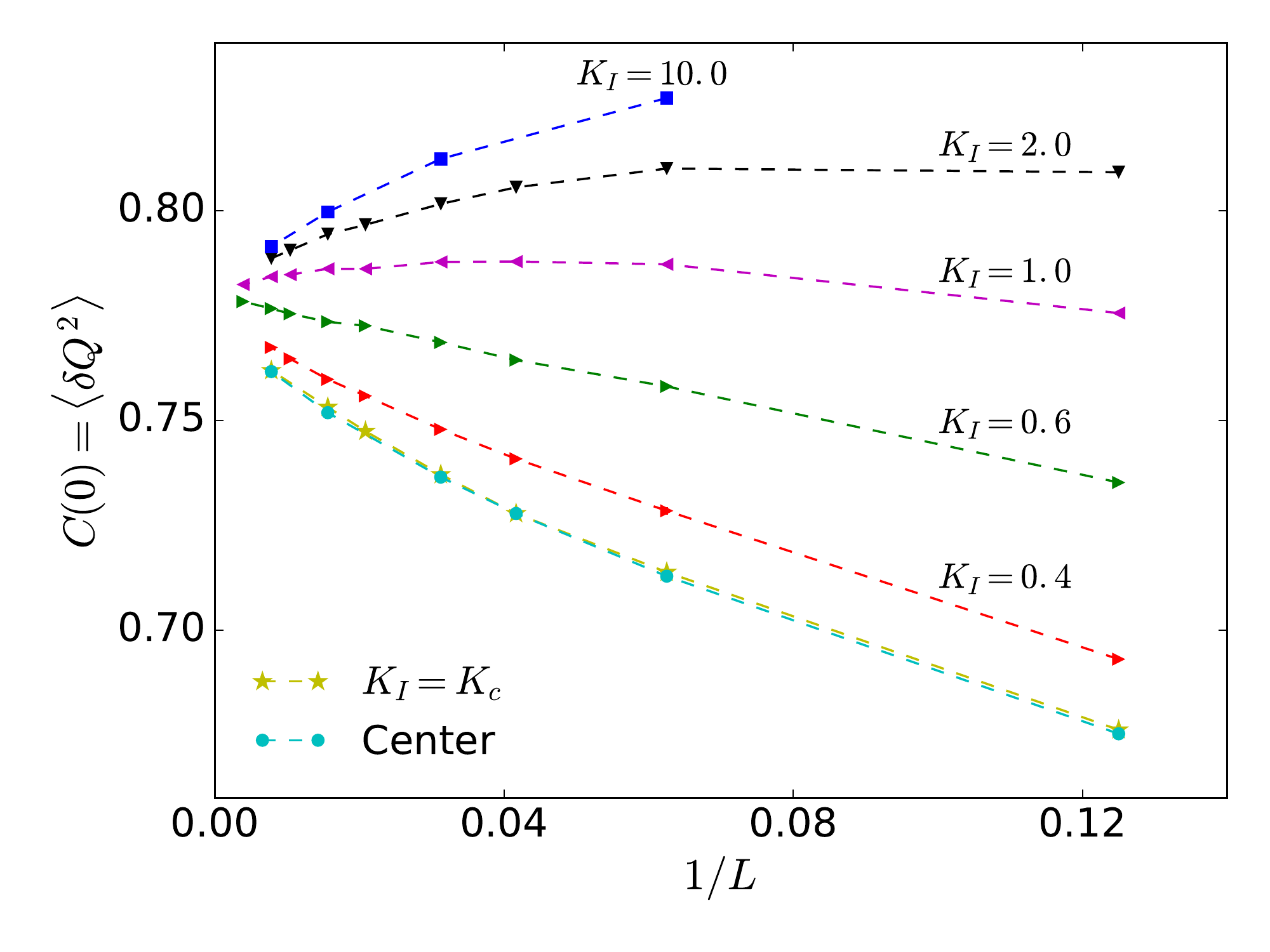}
  \caption{The finite-size  flows of the rescaled charge compressibility at the boundary critical point: the spin impurity model with different coupling strength $K_{I}$ and the static spinless impurity model. All the flows tend to converge to the same universal value  $C(0)=0.780(3)$. }
 \label{fig:N2}
\end{figure}

First, we simulate the J-current model (\ref{J-current}) with a spin impurity at the origin to study the universal function $C(x)$ at $x=0$, i.e., at the critical point $h_z=0$. In Fig.~\ref{fig:N2}, we present the finite-size flows for the rescaled compressibility at the BQCP, which is expected to saturate to a universal constant $C(0)$ in the thermodynamic limit. We compare the flows of the J-current model at different strengths of the impurity-environment interaction $K_{I}$. In all the cases, the flows extrapolate to the same universal constant $C(0)=0.780(3)$, consistent with the idea of sharing  the same universal long-wave theory. 
The opposite trends at $K_{I} < 0.6$ and $K_{I} > 1.0$ indicate that there is an intermediate coupling strength $0.6< K_* <1.0$ at which the leading finite-size correction vanishes. It is thus reasonable to choose $K_I \approx K_*$ to minimize the finite-size systematic error.

We also calculate the flow of the static impurity model (\ref{Jcurrent-center}). Interestingly, the flow is almost identical to that of the spin impurity model with a coupling strength $K_{I}=K_c$, where $K_c$ is the bulk critical coupling strength in Eq.~(\ref{J-current}). 

We now turn to the $h_z$-dependence of the universal compressibility.  In the simulation of the original J-current model (\ref{J-current}), we find that the data for the compressibility suffers from significant finite-size deviations caused by the renormalization of the environment coupling strength $K$ and the impurity-environment coupling strength $K_I$. Therefore, we minimize the renormalization of $K$  by working with the improved J-current model (\ref{improved-J-current}) at $K_I=0.43$, which proves to be the optimal coupling strength for this model.

\begin{figure}[h!]
  \centering
  \includegraphics[width=1.\linewidth]{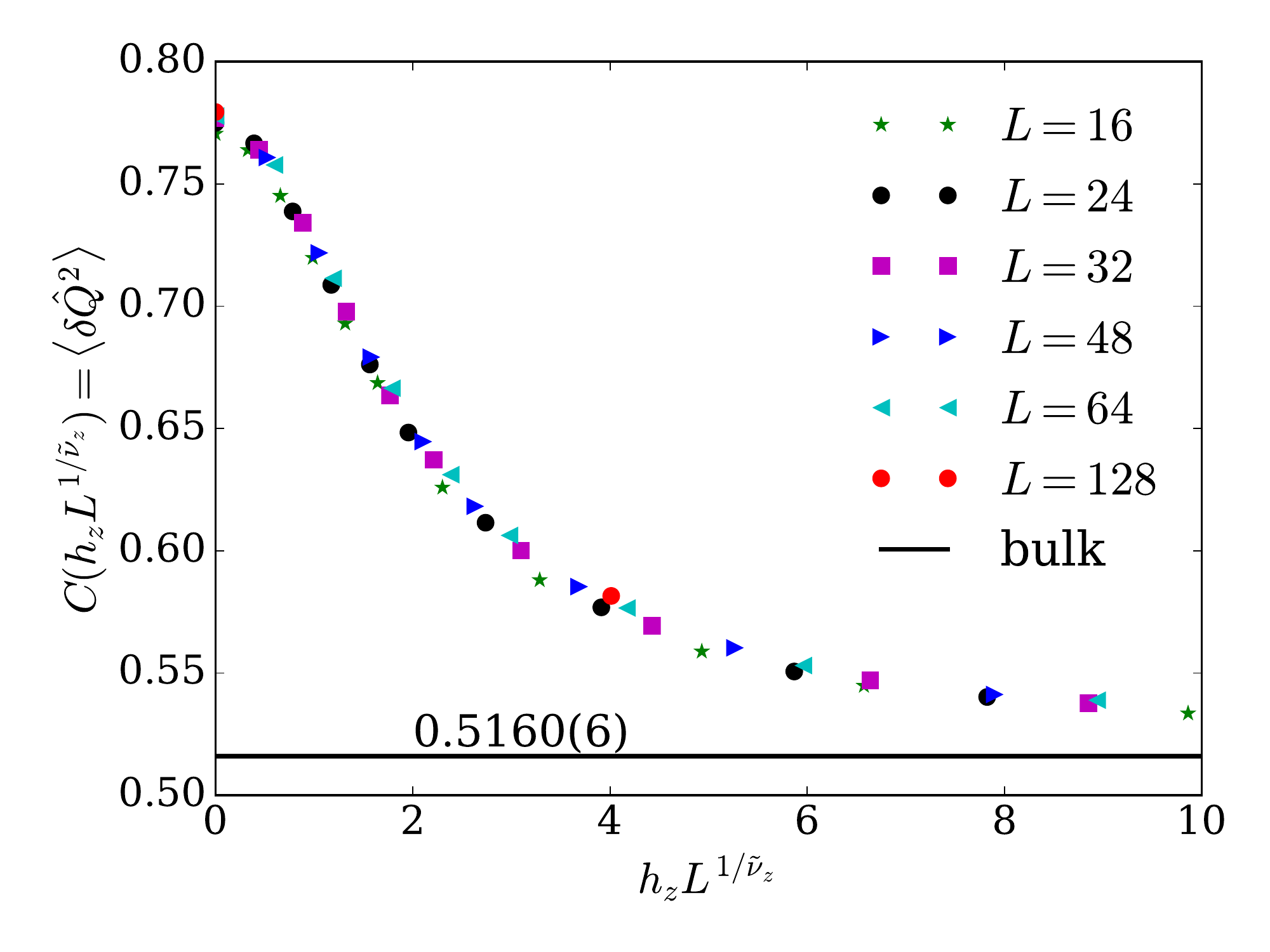}
 \caption{The universal compressibility as a function of the rescaled detuning $h_z L^{1/\nu_z}$ from the boundary critical point. This function is obtained by collapsing the compressibility curves for different system sizes. The critical exponent $\nu_z=2.33$ is used. The black horizontal line shows the universal compressibility of the system in the absence of spin impurity.}
\label{fig:N2_H}
\end{figure}

After rescaling the magnetic field $h_z \rightarrow h_zL^{1/\nu_z}$, we find---see Fig.~\ref{fig:N2_H}---that all the curves---even for relatively small system sizes---collapse nicely by setting the boundary critical exponent to be $\nu_z=2.33$. The master curve is the universal compressibility $C(x)$.  The universal value at the BQCP proves compatible with the universal value $C(0)=0.780(3)$ obtained from the original J-current model (\ref{J-current}). Note also that $C(x)$ eventually saturates to a constant $C(\infty)$ at large $h_z$ and low temperature. The value of $C(\infty)$ is consistent with the universal constant $0.5160(6)$, which is the total compressibility calculated for the bulk quantum critical system when the impurity is absent \cite{conductivity}. We find the effective impurity spin susceptibility at the BQCP to be $[C(0)-C(\infty)]/T=0.264(3)/T$. Despite the fact that the spin impurity is strongly coupled to the environment, the effective spin susceptibility is only slightly different from that of an isolated spin impurity, which is $S(S+1)/3T=1/4T$ (with $S=1/2$). 

\subsection{Impurity Dynamics}

We now turn to the correlators, $\chi_{\alpha}(\tau)$, $\alpha = z,\perp$ [see Eq.~(\ref{eq:corr})], associated with the response of the spin impurity to local magnetic fields $h_z$ and $h_\perp$, respectively. We calculate these correlators with the original J-current model (\ref{J-current}) and an optimized impurity-environment coupling strength $K_I=0.6$.

\begin{figure}[h!]
 \centering
  \includegraphics[width=1.\linewidth]{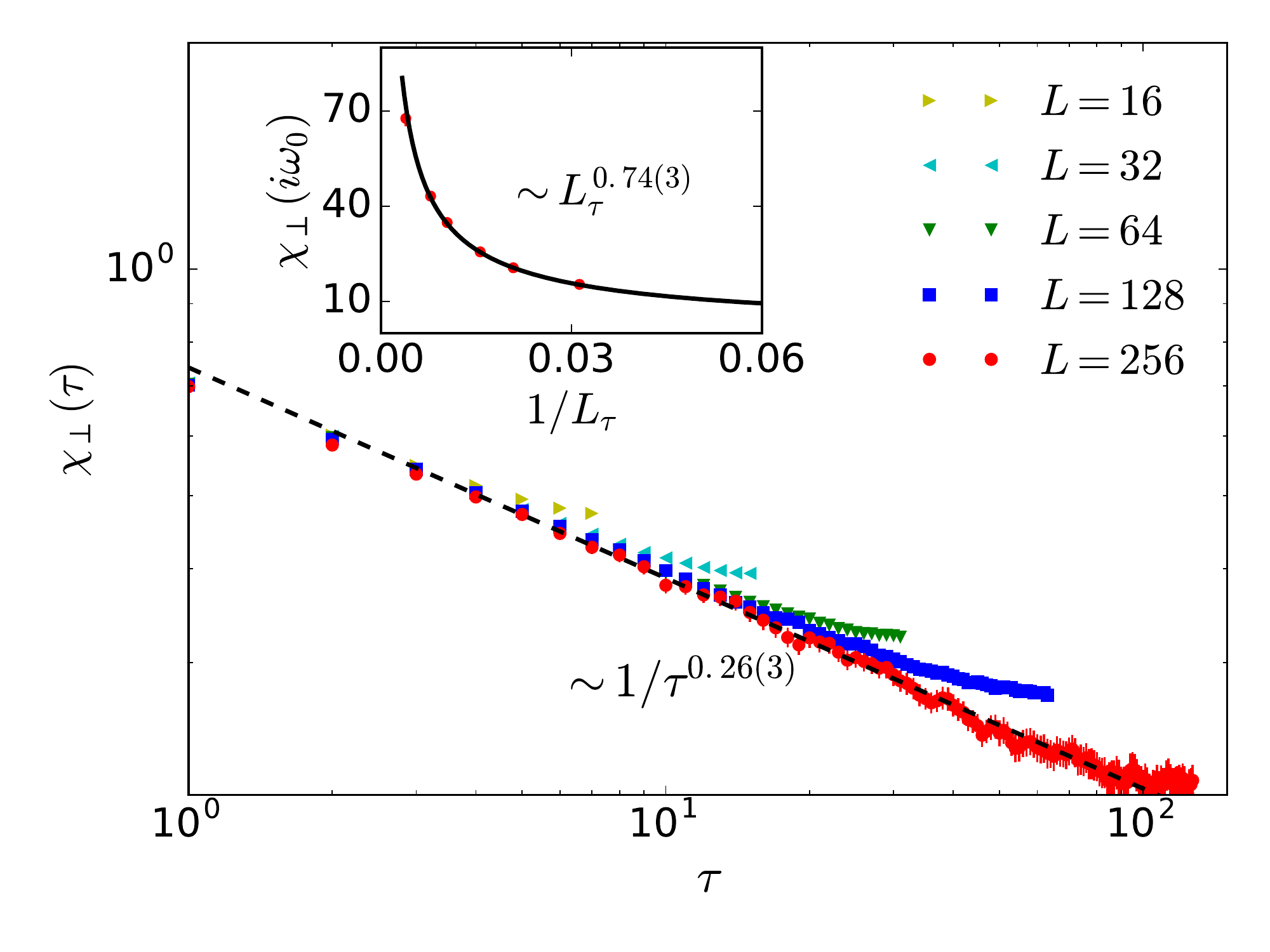}
\caption{The transverse correlation function $\chi_{\perp}(\tau)$ at the BQCP for different system sizes $L$. In the thermodynamic limit $L\to \infty$, there develops a  power-law tail (the dashed line) expected from the general analysis, and yielding $\nu_{\perp}=1.15(3)$. 
The inset shows how this exponent is extracted from the finite-size scaling of the zero-frequency transverse correlation function: $\chi_{\perp}(i\omega_0=0) \sim L_{\tau}^{2/\nu_{\perp}-1}$ (see Sec.~\ref{sec:spin_impurity}).}

\label{fig:Green} 
\end{figure}

 In Fig.~\ref{fig:Green}, we show the transverse correlation function $\chi_{\perp}(\tau)$ at the BQCP for different system sizes. In the thermodynamic limit $L\to \infty$, the asymptotic decay of the correlation function follows the power law. The power-law decay is consistent with Eq.~(\ref{eq:corr_tau0}), which translates the decay exponent $0.26(3)$---obtained by fitting the data---into 
 \begin{equation}
 \nu_{\perp}=1.15(3).
\label{eq:nu_perp}
\end{equation}

\begin{figure}[h!]
  \centering
  \includegraphics[width=1.\linewidth]{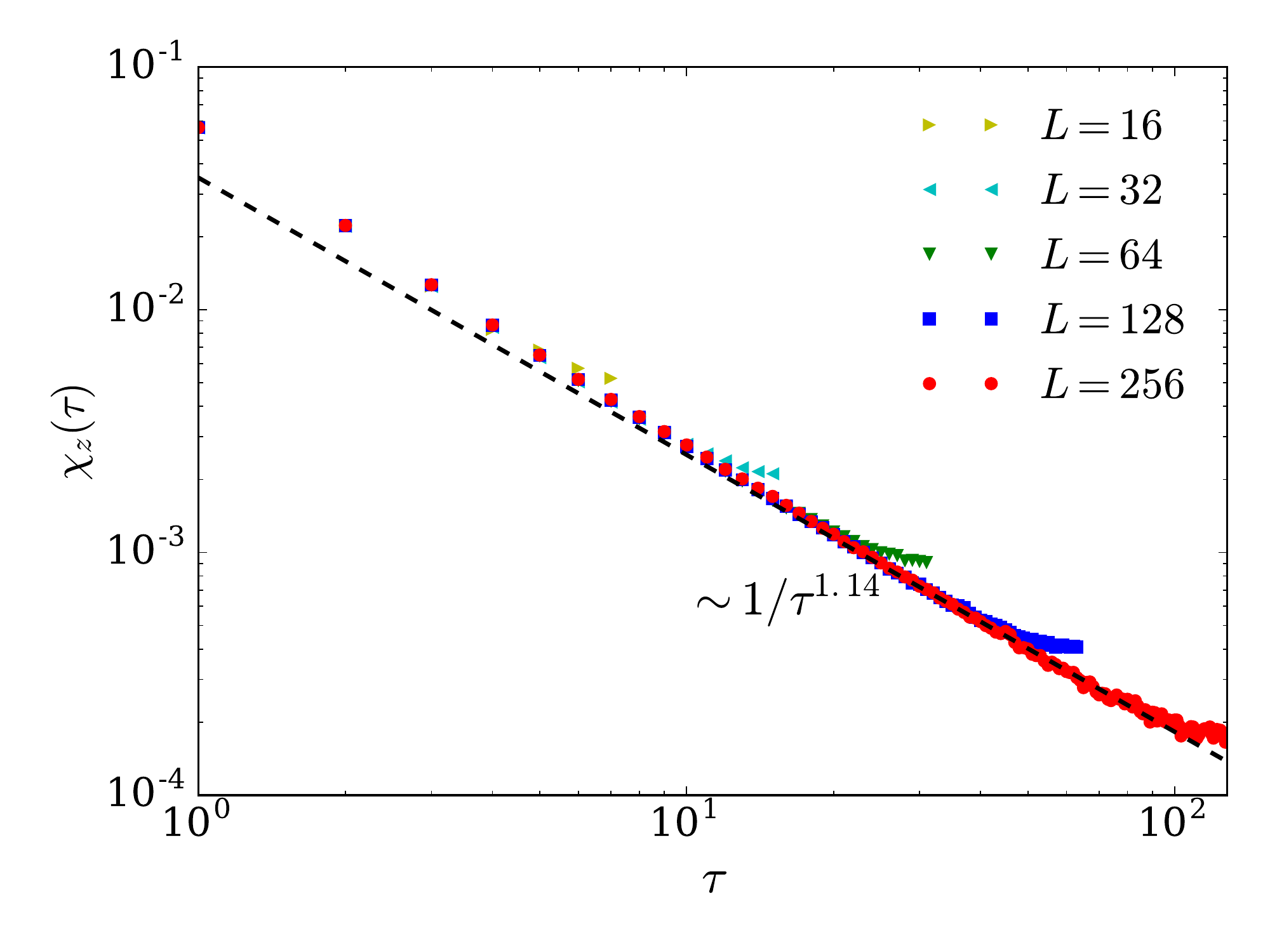}
 \caption{The longitudinal correlation function $\chi_{z}(\tau)$ at the BQCP for different system sizes $L$. In the thermodynamic limit $L\to \infty$, there
develops a  power-law tail (the dashed line) expected from the general analysis (see the text), consistent with $\nu_z=2.33(5)$.}
\label{fig:NNcorr}
\end{figure}

In Fig.~\ref{fig:NNcorr}, we show the longitudinal  correlation function $\chi_{z}(\tau)$ at the BQCP for different system sizes. While this figure is the counterpart
of  Fig.~\ref{fig:Green}, the way we process the thermodynamic-limit power-law tail is different. Since we already know $\nu_z=2.33(5)$, we simply make sure 
that the data is consistent with the prediction of Eq.~(\ref{eq:corr_tau0}), the dashed line.

\begin{figure}[h!]
 \centering
  \includegraphics[width=1.\linewidth]{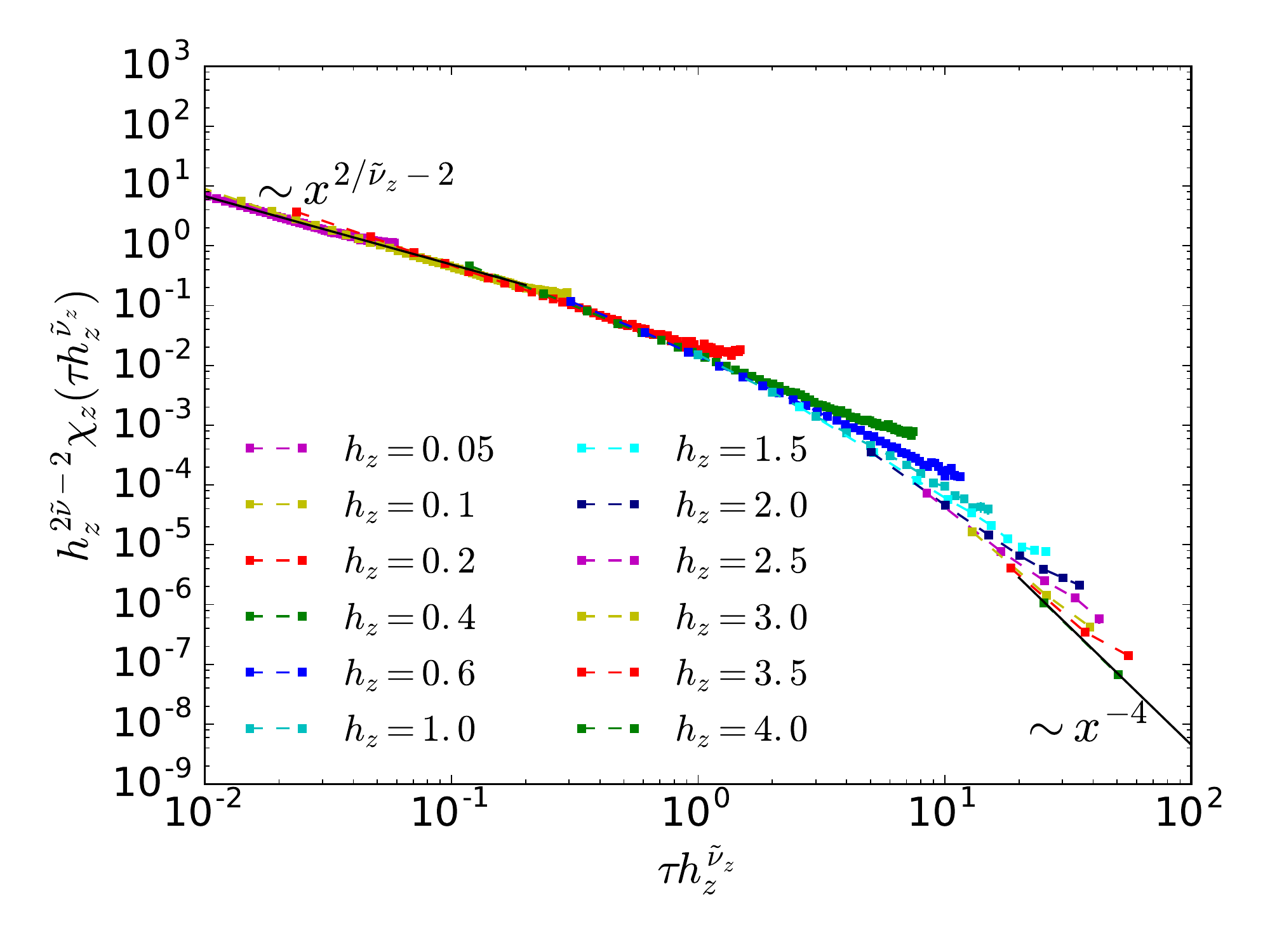}
\caption{\label{fig:corr_z} The data collapse for the function functions $\chi_z(\tau)$ at different values of the magnetic field $h_z$ at a given system size $L=128$. Both axes are properly rescaled so that the universal parts of the correlation functions match each other thus yielding the master curve. This curve starts with a boundary critical behavior $x^{2/\nu_z-2}$ with $\nu_z=2.33(5)$ at small $x$ and eventually evoles into asymptotic behavior $x^{-4}$ at large $x$.}
\end{figure}

Once moving away from the BQCP with a small magnetic field $h_z$, a large correlation time $\xi_z \sim |h_z|^{-\nu_z}$ emerges. 
The universal correlation function $\Psi_z(x)$ [see Eq.~(\ref{eq:corr_tau})] behaves very differently in  the short- and long-time limits. The short-time behavior within the correlation time coincides with the universal correlation function right at the BQCP, as in Fig.~\ref{fig:NNcorr}, while the long-time behavior is expected to be $\sim x^{-4}$. Due to the relatively large exponent, the long-time physics turns out to be extremely difficult to resolve within a single simulation box. We overcome this subtlety by using the flowgram method \cite{kuklov, dcp}; see  Fig.~\ref{fig:corr_z}. First, we calculate various longitudinal correlation functions for different magnetic fields, then rescale $\tau \rightarrow \tau h_z^{\nu_z}$ and $\chi_z(\tau) \rightarrow \chi_z(\tau) h_z^{2\nu_z-1}$ to match the universal parts of different correlation functions. We observe an excellent collapse of the data onto a single master curve yielding the universal function $\Psi_z(x)$, covering both the short- and long-time scales. From Fig.~\ref{fig:corr_z}, we clearly see that the universal function starts with $x^{2/\nu_z-2}$ at small $x$ and eventually decays as $x^{-4}$ at large $x$. The challenge of bruit-force observation of the long-time tail is clearly seen from this figure. Taking $h_z=1.0$ as an example, the tail starts to dominate beyond the time scale $\sim 20$. However, the amplitude of the correlation function at this time scale drops to  $~10^{-6}$, which is already comparable to the typical statistical error in our Monte Carlo simulations. 

%

\section {O(3) Halon in a Lattice Spin System}
\label{sec:O_3}

While the O(N) quantum rotor model captures all the universal properties of the O(N) halon, it is  important---from the experimental viewpoint---to discuss 
real-world quantum models as well. So far, we have done that for O(2) halon only. This section deals with an experimentally relevant implementation of the O(3) halon in a dimerized Heisenberg antiferromagnet.

\begin{figure}[h!]
 \centering
 \includegraphics[width=0.3\textwidth]{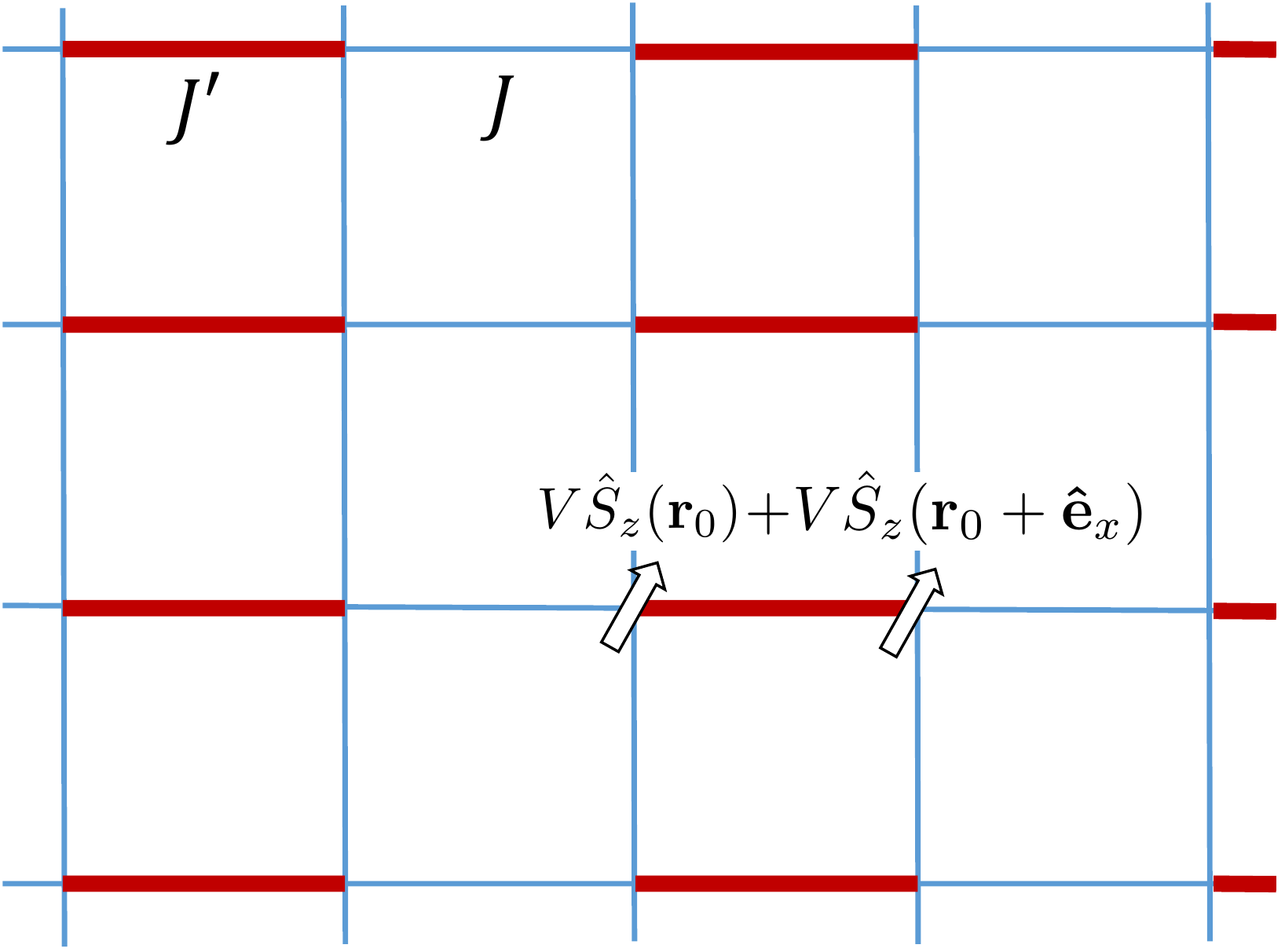}%
 \caption{\label{fig:spin_model} A composite local-magnetic-field impurity in a dimerized Heisenberg antiferromagnet. The environment is a spin system on a coupled-dimer lattice. The exchange couplings are $J'$ for the solid red bonds and $J$ for the blue bonds. The impurity is a local magnetic field coupled to the spin-z projection of a pair of spins in the same unit cell.}
 \end{figure}

Consider a square-lattice spin model with two different coupling strengths,
\begin{equation}
\label{eq:dimer}
\hat{H}_{\rm AF}=J \sum_{\langle i,j \rangle} \hat{\bf{S}}_i \cdot \hat{\bf{S}}_j+J' \sum_{\langle i,j \rangle'} \hat{\bf{S}}_i \cdot \hat{\bf{S}}_j,
\end{equation}
where $\hat{\bf{S}}_i$ are spin-$1/2$ operators on lattice sites, and the exchange coupling between different sites form a coupled-dimer pattern as shown in Fig.~\ref{fig:spin_model}. The dimensionless coupling $J'/J$ controls the ground-state phase diagram of the system. When the intra-dimer coupling $J'$ dominates, the ground state is a paramagnet consisting of an array of spin-singlets. On the other hand, when the inter-dimer coupling $J$ dominates, the ground state is an antiferromagnet where neighboring spins point in the opposite directions. A continuous quantum phase transition happens at $J'/J=1.9096(2)$ \cite{dimer}, and the infrared physics of this critical point is described by a $(2+1)$-dimensional O(3) critical theory. 

This model has wide applications in condensed matter physics. It describes dimerized antiferromagnetic materials like $\textit{TlCuCl}_3$. It is also relevant to the effective physics of many exotic electron systems, such as the $CuO_2$ layers of the cuprate superconductors \cite{superconductor} and double-layered integer quantum Hall systems \cite{hall}. The model can also be realized with ultracold fermions in 2D optical double-well superlattice potential \cite{fermi_hubbard1, fermi_hubbard2, fermi_hubbard3}. 

It is known \cite{Sachdev_book} that the long-wave properties of the dimerized antiferromagnet are captured by the O(3) quantum rotor model (\ref{eq:rotor}). The rotor is defined for each unit cell and can be considered as an effective representation of a pair of antiferromagnetically coupled spins. The rotor orientation operator ${\bf{\hat{n}}}_i$ corresponds to the local staggered magnetization ${\bf \hat{S}}_i-{\bf \hat{S}}_j$ where $i,j$ are sites in the same unit cell. The rotor angular momentum operator ${\bf{\hat{L}}}_i$, which plays the role of the Noether charge density, corresponds to the local uniform magnetization in each unit cell ${\bf \hat{S}}_i+{\bf \hat{S}}_j$.

Given the above-described mapping, a counterpart of the impurity Hamiltonian Eq.~(\ref{eq:rotor_imp}) can be implemented as
\begin{equation}
\label{eq:AF_impurity}
\hat{H}_{\rm imp}= V [\hat{S}_z({\bf {r}}_0)+ \hat{S}_z({\bf {r}}_0+{\bf{\hat{e}}}_x)] .
\end{equation}
Here the impurity is represented by the local magnetic field $V$ coupled to two spins in the same unit cell. When $V$ reaches the critical value $V_c$, the singlet state in the unit cell  breaks down to a direct-product state, causing the total charge $Q$ to abruptly change by one spin-projection quantum. 

\begin{figure}[h!]
 \centering
  \includegraphics[width=1.\linewidth]{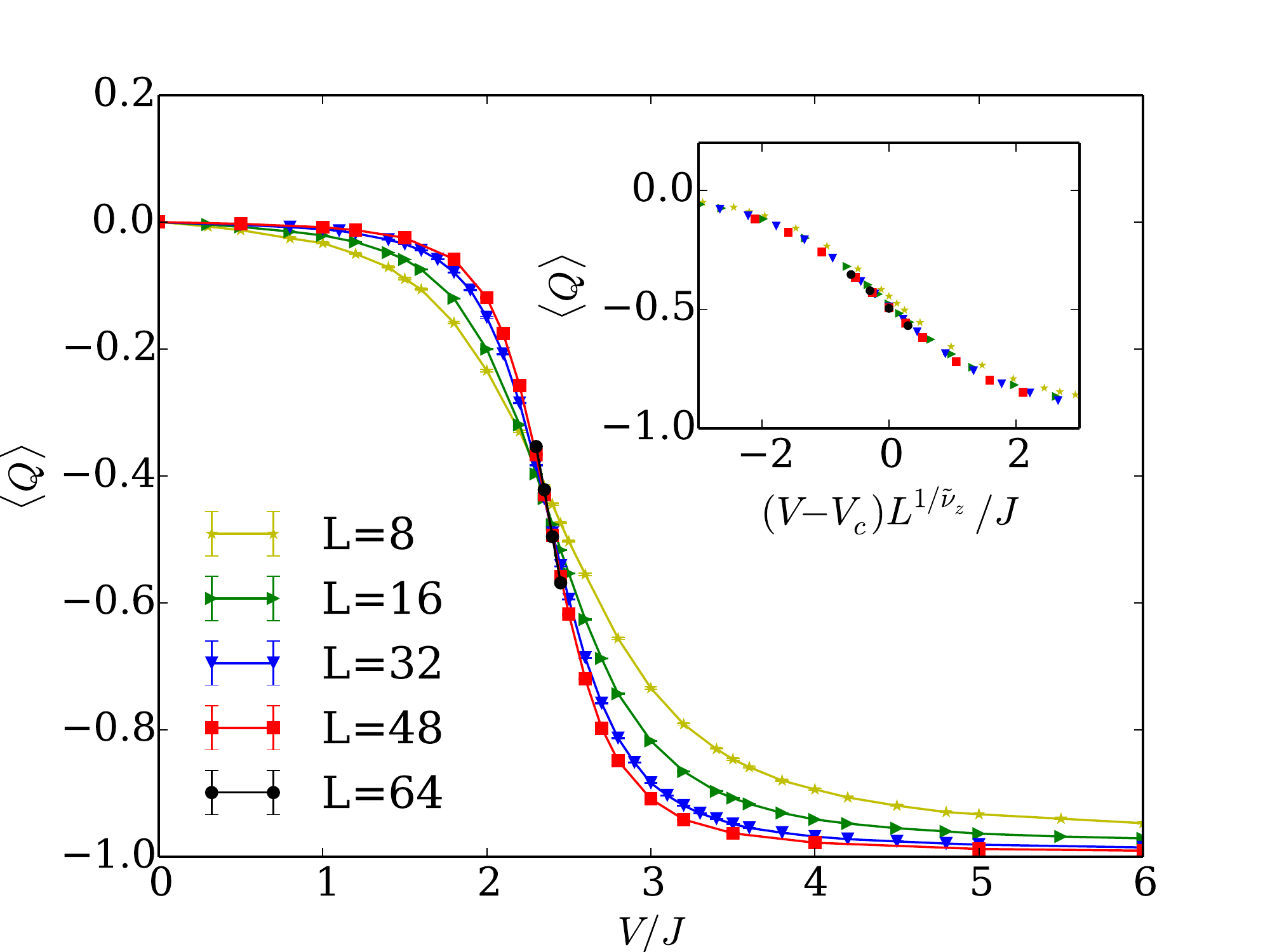}
\caption{\label{fig:lattice_spin} The total uniform magnetization in the dimerized Heisenberg model as a function of the (rescaled in the inset) strength of the local field at different system sizes. The simulation is performed in the grand-canonical ensemble.}
\end{figure}

On the quantitative side, we perform large-scale  simulations of the model (\ref{eq:dimer})--(\ref{eq:AF_impurity}) by the worm algorithm. In Fig.~\ref{fig:lattice_spin}, we show the data---both raw and rescaled---for the total charge $Q$ as a function of $V$. With the data collapse upon rescaling (see the inset), we find  $V_c/J=2.40(2)$ and  
$\nu_z=2.32(8)$. Similar to its O(2) counterpart, the exponent $\nu_z$ is larger than $2$, and  the general analysis of Sec.~\ref{sec:spin_impurity} applies. 

Let us briefly discuss the Bose-Kondo-impurity counterpart of the coupling (\ref{eq:AF_impurity}). While being  less relevant to the experiment, the model is interesting on its own. The key observation is that the degeneracy between the singlet/direct-product states at BQCP can be effectively described by a pseudo-spin-$1/2$ degree of freedom. The Bose Kondo model is then introduced by replacing two physical spins in the same unit cell with one pseudo-$1/2$-spin, and coupling the pseudo-spin to the neighboring physical spins via an XY-type interaction,
\begin{equation}
\label{eq:spin_imp_AF}
\hat{H}_{\rm AF} \to \hat{H}_{\rm AF}+ \gamma \sum_{n} \left[{\hat S}_ +  \,{\hat S}_-({\bf {r}}_n)\, + \,  {\rm H.c.} \right]+h_z\hat{S}_z .
\end{equation}
Here the sum runs over all the nearest neighbors of the unit cell replaced with the pseudo impurity. One advantage of working with this model is that the BQCP is exactly known to be $h_z=0$ by the explicit spin-flip symmetry.

\section{Conclusions and Outlook}
\label{sec:conclusion}

We have discussed, both qualitatively and quantitatively, the physics of a (static) halon---a special fine-tuned state of a (static) impurity in a gapless and incompressible 2D O($N$), $N\ge 2$, quantum-critical environment, with a special emphasis on the O(2) case and also paying a considerable attention to the O(3) case. While generically supporting charge quantization, the environment  allows for a non-trivial charge fractionalization phenomenon---the halon effect---on approach to the boundary quantum critical point (BQCP), at which the charge of the impurity changes  by $\pm 1$. On approach to the BQCP, the well-defined integer charge carried by the impurity gets fractionalized into two parts: a microscopic core with half-integer charge and a critically large halo carrying a complementary charge of $\pm 1/2$.

The microscopic details of the impurity can be rather different, provided the following two necessary conditions are satisfied: (i) the impurity couples to the Noether's charge density associated
with this or that U(1) symmetry in such a way that the coupling respects the symmetry and thus the conservation of the total U(1) charge; (ii) the quantum-critical environment features the (emergent) particle-hole symmetry with respect to the U(1) charge. In a general case, the U(1) group in question is a subgroup of a larger symmetry group, the latter being respected by the quantum-critical
environment, but not the impurity.

As observed independently by Whitsitt and Sachdev \cite{Whitsitt_Sachdev} and us, the universality class of halon BQCP is captured by a Bose Kondo model of pseudo-spin-$1/2$ impurity coupled to the quantum-critical environment, in such a way that the rotational symmetry in the pseudo-spin $xy$-plane is respected, with a small local ``magnetic" field  along the pseudo-spin $z$-axis playing the role of control parameter driving the system away from the BQCP. On the approach to BQCP, the half-integer projection of the pseudo-spin on its $z$-axis gets delocalized into a halo of critically divergent radius, capturing the essence of the phenomenon of charge fractionalization. 

The Bose Kondo model sheds an extra light on the universal properties of halon, characterized by two independent boundary critical exponents. One exponent, $\nu_z$, controls the impurity charge dynamics (or the longitudinal dynamics in the Bose Kondo model), another one, $\nu_\perp$, controls the order-parameter field dynamics (or the transverse dynamics in the Bose Kondo model) near the impurity site. The response functions related to these two cases are qualitative different. The exponent $\nu_z$ also controls the universal and nontrivial (featuring integrable singularity) charge density profile of the halon at short distances.

By changing the pattern of global charge fluctuations, the halon has a significant effect on the total compressibility of the system---comparable with the contribution of the rest of the system. As a result, the critical total compressibility increases by a nontrivial universal constant. 

Our quantitative study of halons in O(2) and O(3) quantum-critical 2D systems is based on large-scale worm-algorithm Monte Carlo simulations. In the O(2) case, we simulated effective classical models for both the trapping center problem and its Bose Kondo counterpart. We verified various universal scaling relations and quantified the universal features. In particular, we found the two independent critical exponents to be $\nu_z=2.33(5)$ and $\nu_{\perp}=1.15(3)$. We also simulated a lattice spin system of  O(3) quantum criticality and demonstrated that a local-magnetic-field impurity (counterpart of the trapping center) does lead to a halon BQCP. Here the critical exponents $\nu_z$ is found to be $2.32(8)$, which is very close to the O(2) case. The next-leading order $\epsilon$-expansion calculations performed by Whitsitt and Sachdev \cite{Whitsitt_Sachdev} yield $\nu_z\approx 2.66$ and $\nu_{\perp}\approx 1.08$ for O(2) case and  $\nu_z\approx2.78$ for O(3) case. Those results agree quite well with our first-principle data. 

Ultracold bosonic atoms in optical lattices ~\cite{bloch_review}---being a nearly ideal realization of the Bose-Hubbard model  (\ref{BH})---provide a natural experimental context for the  halon effect. In view of the mapping of the halon physics onto that of the Bose Kondo model, one can also view this experimental setup as an implementation of the Bose Kondo model.
The trapping center (a potential bump or well) can be introduced by applying a tightly focused laser beam, perpendicular to the plane of the 2D lattice. Thanks to recent advances in single-site techniques~\cite{imaging, holographic}, the trapping center can be rendered as small as a single site, thus  directly implementing the impurity Hamiltonian (\ref{BH}). 

Obvious experimental challenges are the uniformity of the system,  low temperature, and  small deviation from the unity (or any other integer) filling of the lattice.
Each of the three circumstances leads to systematic errors imposing an upper bound, $r_{\rm syst}$, on the halo size $r_0$, the universal halon regime taking place only at $r_0 \ll r_{\rm syst}$.

For enhanced control of the quantum criticality in the bulk as well as for accurately tuning the trapping center to the BQCP, the experiment can take advantage of the data obtained
from first-principles numeric simulations of the Hamiltonian (\ref{BH}). 
The most desired experimental observables are the ones revealing the static and dynamic properties of the halon discussed in Sec.~\ref{sec:spin_impurity}. Of special interest is the halon dynamics that cannot be addressed by unbiased numerical methods. In view of the mapping onto the Bose Kondo model,  the study of the halon dynamics can shed a certain light 
on the impurity dynamics in some real materials ~\cite{bkondo3,bkondo4}. 

For the future study of the halon physics, the following two circumstances seem very intriguing.

(i) The fact of fractionalization of the charge implies that the charge of the core and the halo charge are entangled. Diverging radius of the halo thus results in the long-range entanglement.
Going from a static singe halon to mobile halons of finite concentration then leads to an extremely nontrivial system with most exotic (entanglement-assisted) long-range interaction.

(ii) In recent years, it has been established that in two spatial dimensions, the superfluid near the quantum critical point features the critically defined massive-Goldstone mode (also known as the Higgs mode)  causing a sharp resonance peak in  relevant spectral functions \cite{pod11, higgs, higgs_nikolay, pod12, higgs_kun1, Snir1, Rancon, higgs_kun2}. There is also numerical evidence showing that a massive-Goldstone-like mode may also exist in the critical Mott-insualtor or even in the critical liquid regime\cite{higgs_nikolay, higgs_kun1, Rancon}. It would be
interesting to understand how/whether the halon (or the quasi-halon in the critical superfluid and the critical Mott-insulator regimes) is coupled to the massive-Goldstone mode: Can one see the massive-Goldstone signal in the response function of the trapped charge? 

\section{Acknowledgements}

The authors are grateful to Nikolay Prokof'ev, Subir Sachdev, Seth Whitsitt, Immanuel Bloch, Markus Greiner, Richard Schmidt, Zhen-Sheng Yuan, Yashar Komijani for valuable discussions. This work was supported by the National Science Foundation under the grant DMR-1720465 and the MURI Program ``New Quantum Phases of Matter" from AFOSR. YD acknowledges  the Ministry of Science and Technology of China (under Grant No. 2016YFA0301604) and the National Natural Science Foundation of China (under Grant  No. 11625522).

\end{document}